\documentclass[letterpaper]{aa}
\usepackage{epsfig}
\usepackage{amsmath}
\usepackage{txfonts}
\usepackage{amssymb}
\begin{document}

\title{Gravitational effects on the high energy emission of accreting black holes}

\author{T. Suebsuwong, J. Malzac, E. Jourdain \and A. Marcowith}
\authorrunning{T. Suebsuwong et al.}

\institute{Centre d'\'Etude Spatiale des Rayonnements, CNRS-UPS, 9 avenue du Colonel Roche, 31028 Toulouse Cedex 4, France
 \\ \email{suebsuwo@cesr.fr}}

\date{}

\abstract
{{We extend the investigation of general relativistic effects on the observed X-ray continuum  of Kerr black holes in the context of the light bending model (Miniutti \& Fabian, 2004).}}
{ Assuming a ring-like illuminating source, co-rotating with the underlying accretion disk, we study the shape
 and normalisation of the primary and disc reflected continuum as well as the dependence of the observed spectrum
  on the line of sight for various source heights and radii. }
{ These calculations are performed using Monte-Carlo methods to compute
 the angle dependent reflection spectrum from the disc.
The effects of general relativity are illustrated by a comparison
 with Newtonian and Special Relativity calculations.}
 { Relativistic distortions can strongly affect the shape of
  the reflected spectrum. Light bending can dramatically increase
  the observable reflected flux and reduce the primary emission.
 In addition, multiple reflections due to the reflected photons deflected
toward the disc can alter significantly the shape of the spectrum above 10 keV.
 We explore the predicted variations of
 the observed reflected and primary fluxes with the height and radius of the source.
 Large variations of the ring radius at constant height can lead to
an (unobserved) anti-correlation between primary and reflected flux.
In another side, the variability behaviour of several sources can be reproduced
 if the ring source radius is
small ($<5\,r_{\rm g}$), and its height varies by a large factor.
In particular, a non-linear
 flux-flux relation, similar to that observed in several sources, can be produced.
 We compare our model with the flux-flux plot of NGC4051, and
 find an agreement for low inclination angles ($<20^o$),
  ring source radius $\lesssim$ 3 $r_{\rm g}$
  and a height varying between 0.5 to 10 $r_{\rm g}$ .
 Regarding the angular distribution of the radiation, we
find some important qualitative differences with respect to the Newtonian case.
 The reflected flux at larger inclination is relatively stronger than in the
 Newtonian model, the reflection fraction increasing with inclination.}
{}
\maketitle
\section{Introduction}
The power-law X-ray spectra of Seyfert galaxies and black hole binaries
are generally believed to originate from the central parts of the accretion flow
trough inverse Compton process in a relativistic plasma.
This hot plasma could be either the innermost part of a hot geometrically
thick optically thin accretion disc (Shapiro et al. 1976; Narayan \& Yi 1994)
or a hot magnetic corona atop a cold standard geometrically thin optically
thick accretion disc (Bisnovatyi-Kogan \& Blinnikov 1976; Haardt \& Maraschi 1993).
In addition, the high energy spectra often show the presence of reprocessing features
such as a  $Fe$ $K\alpha$ line  at (or close to) 6.4 keV and a Compton reflection bump
peaking around 30 keV. Such features are expected from reprocessing and reflection of the hard X-ray continuum
by cold and dense gas in the accretion disc (George \& Fabian 1991; Matt et al. 1991)
and/or by more distant material (Malzac \& Petrucci 2002), such as the torus
of the unified model for AGN (Antonucci 1993). In many instances, there is strong evidence
that these reflection features are formed in the accretion disc in the direct vicinity
of the black hole.  Indeed during the last decade, many studies have shown that this
radiation exhibits the imprints of strong gravitational field
and rapid orbital motion of matter near a black hole (Fabian et al. 1989; Laor 1991; Matt et al. 1996; Reynolds \& Begelman 1997;
Martocchia et al. 2000).
The spectacular relativistic profile of the iron line observed by ASCA in the Seyfert galaxy
MGC6-30-15 (Tanaka et al. 1995) confirmed by BeppoSAX (Guainazzi et al. 1999),
and XMM-Newton (Wilms et al. 2001) and Chandra (Lee et al. 2002) is the most impressive example.
Although recent Chandra and XMM results have shown that the presence
of relativistic lines in Seyfert 1s is far from being the rule,
 broad and asymmetric profiles appear to be quite common (Fabian 2004).
Gravitational effects  affect not only the line profile but also the shape and strength
of the reflected continuum.
Actually, the spectra of many several luminous Seyfert galaxies are
very well described by photoionized  and strongly relativistically blurred reflection models
(Fabian et al. 2004; Crummy et al. 2005, Fabian et al. 2005; Porquet 2006).
In these sources the primary continuum often appears to be strongly suppressed
which could represent further evidence for light bending effects.
In a number of Narrow Line Seyfert 1 (NLS1) galaxies and galactic black holes
in the very high state, the variabilities of the continuum and of the iron line
are decoupled,  in apparent contradiction with the predictions of simple
disc reflection models (see, e.g. Miniutti et al. 2003; Miniutti, Fabian \& Miller 2004;
 Fabian et al. 2004; Rossi et al. 2005).

 Miniutti \& Fabian (2004) interpret these
 otherwise puzzling variability properties in terms of light bending effects (previously suggested by Fabian \& Vaughan, 2003).
In their model, the active coronal region(s)  illuminating the disc
are idealised as a ring source at some height above, and corotating with
the accretion disc. When the source is close enough to the black hole, the primary
 component  is strongly suppressed leading to reflection dominated spectra.
 Moreover, as shown by these authors, fluctuations in the height of the source
 can lead to strong variability in the primary component
 with little variability in the reflected flux, as observed.

In this work we investigate further the light bending model of Miniutti and Fabian.
We use detailed Monte-Carlo simulations to compute not only the iron
line profile (as done by these authors) but also the shape of the reflection continuum.
Previous calculations of the relativistically blurred reflection spectra
are based on a simple convolution of a Newtonian reflection with a relativistic kernel
assuming a given emissivity law in the accretion disc (e.g. Dov\v ciak et al. 2004;
Ross \& Fabian 2005). In these latter works the authors focus only on the shape of the reflected spectrum while very little attention
 is paid to the effects on the primary emission and how they affect the ratio between
  primary and reflected fluxes.
Moreover in all the previous studies of the GR effects on the iron line and reflection
continuum, the possible multiple disc reflections, due to the reflected photons being gravitationally deflected again
  toward the disc, were neglected.
  Here, we will compute consistently the reflection and iron line expected
from the ring source model including the contribution from multiple reflections and
 taking into account the exact angular dependence
of the illuminating radiation in Kerr geometry. We will present a detailed  investigation
of the effects of general relativity
on both the shape and normalisation of the primary and reflected components,
with a particular emphasis on their dependence on the inclination of the line of sight.

We also extend the work of Miniutti \& Fabian by showing
the effects of varying the radius of the ring source in addition to its height.
For reference, we also compare all
our general relativistic simulations with the results obtained for the
Newtonian geometry with both a static and a rotating disc.
This enables us to disentangle the effects of general relativity from
Newtonian and disc rotation effects. In particular, we show that due to light bending,
the changes in the appearance of a source according to the viewing angle
are very different from what is expected from the Newtonian calculations.
Finally our self-consistent computations of the reflected continuum enable us to estimate
the relative variations of the flux in two different energy
bands when the geometry of the ring source changes.
Comparing the results with data, we show that the  predictions of the
light bending model
 would be compatible with the non-linear
flux-flux relation observed in the Seyfert galaxy NGC4051.

The structure of the paper is the following:  Sec.~\ref{sec:modeldescript}
is devoted to the description of our calculations. Sec.~\ref{sec:onaxis}
presents the results for the limiting case in which the
ring source radius is zero i.e. a on-axis point source (the so-called lamp-post model)
and studies the effect of the height of the source. The more general ring source model
is studied in Sec.~\ref{sec:offaxis} and spectra for various source radii
are presented and discussed. In Sec.~\ref{sec:observations} we discuss the relation
between our results and the observations.
We conclude in Sec.~\ref{sec:conclusion}
with a short summary of our results.

\section{Model description}
\label{sec:modeldescript}
\subsection{Model assumptions}
We use the Kerr metric to represent space-time curved by an uncharged rotating black hole. In Kerr geometry,
the black hole is characterised by its mass, $M$ and its angular momentum, $J$ or a dimensionless spin parameter $ a = {Jc}/{GM^2}$.
We consider an extremely rotating black hole with $a=0.998$ (Thorne 1974). The unit of length in this study is the gravitational radius, $r_g = {GM}/{c^2}$.
The inclination of the line of sight to the black hole rotation axis is given by $\theta_{obs}$, we note $\mu_{obs} = cos(\theta_{obs})$.
The hard X-ray primary source has a ring-like axisymmetric geometry and is located above the disc with a radius, $\rho_s$ and a height $h$.
As we consider that the primary source represents a corona close to the disc or flares associated with magnetic activity of the disc,
we can assume that the ring is in co-rotation with the underlying accretion disc.
Since the orbital time-scale in the vicinity of black hole is usually much shorter than the integration time needed in observations, the choice of an axisymmetric geometry is then relevant even if the actual shape of the corona is more complex.
It emits isotropically in its rest frame with an e-folded power law spectrum.
We assume that the accretion disc lies on the equatorial plane of the Kerr geometry, $\theta = {\pi}/{2}$.
The disc is geometrically thin, optically thick and its innermost radius is the marginally stable orbit, $r_{in} = r_{ms}$ extending to an outer radius of $r_{out}=100$.
For $a = 0.998, r_{ms} = 1.23$ (Bardeen et al. 1972). The disc rotates with an angular velocity (Bardeen et al. 1972),
\begin{equation}
 \Omega(r) = 1/(a+r^{3/2}).
 \label{eq:omder}
\end{equation}
The accretion disc material is assumed to be neutral and cold, $T\leq10^6 K$ (George \& Fabian 1991).

The intrinsic spectra emitted by the disc and the ring source are altered by Doppler effects due to the rotation, and the gravitational shift and light bending.
In order to understand and disentangle these different effects,
it is interesting to compare the results of the full general relativistic calculations with simplified approaches.
Namely, we will compare the three following models:
\begin{itemize}
\item Newtonian Model (hereafter NN): a static accretion disc and ring source in Newtonian geometry.
The direction and the energy of the photons in the observer frame are identical to the emitted ones.
\item Special Relativistic Model (hereafter SR): a rotating accretion disc (and primary source) in Newtonian geometry.
The direction and the energy of the photons are affected by beaming and Doppler effects.
\footnote{In this model the radial dependence of the disc velocity is taken to be that of a general relativistic accretion disc
 (Eq.~\ref{eq:omder}). We consider Newtonian geometry for light propagation only.}
\item General Relativistic Model (hereafter GR): a rotating accretion disc in Kerr geometry. Each photon trajectory in this model follows a geodesic in Kerr geometry.
 Its energy is different in the observer and the emitter rest frame due to the gravitational shift.
\end{itemize}

\subsection{Calculation}
We use the Boyer-Lindquist coordinates ($t$, $r$, $\theta$, $\phi$; the black hole rotates in the $\phi$ direction)  as coordinate system of observers at infinity in Kerr geometry.
The static limit is a surface described by $ r_{erg} = M+\sqrt{M^2-a^2cos^2(\theta)}$.
The region of space-time between the horizon and the static limit is called "ergosphere". In this region, all observers must orbit the black hole in the same direction.
We calculate photon trajectories using Carter's equations of motion (Carter 1968).
The constants of motion are derived from the emitted angle and energy in the source rest frame as detailed in Appendix A.

To compute the overall spectrum we use a Monte-Carlo method consisting in tracking the path and interactions of individual photons.
First, we draw the direction and energy of the primary photons. Their directions are drawn from an isotropic distribution in the ring source rest frame.
Their energies are generated from the following energy distribution:
\begin{equation}
 N(E) = N_o\,E^{-\Gamma}\,\rm{exp}\,(\,-\frac{E}{E_{\rm c}}\,)
\end{equation}
where $\Gamma$ is photon index and $E_c$ is cut-off energy.
In most of our simulations and unless specified otherwise, we set $\Gamma=2$ and $E_{\rm c}=$200 keV.

From these quantities we compute the constants of motion and solve the equations of motion. A photon geodesic can either intercept the disc or reach infinity directly. In both cases, we calculate the local power (i.e. either at infinity or on the disc), by
\begin{equation}
\frac{dE_{obs}}{dt_{obs}} = {g^2_{obs}}\, \frac{dE_{em}}{dt_{em}}
\end{equation}
 where $g_{obs} = E_{obs}/E_{em}$, $E_{obs}$ is the energy locally measured
  by observers and $E_{em}$ is the emitted energy.
 The direction of propagation in the local frame can then be derived by inverting Eqs~\ref{eq:contantsofmotion1} -~\ref{eq:contantsofmotion3}.

 If a photon does not intercept the disc, its energy and direction at infinity are stored and used to build up the observed spectrum.
Otherwise, we use the Monte-Carlo code of Malzac et al. (1998) to track its interactions inside the accretion disc.
This reflection code takes into account Compton diffusions in cold neutral matter, photoabsorption and iron fluorescence.
The column density of the disc is fixed at $10^{26}$ $\rm cm^{-2}$. We use the photo-absorption opacities from Morrisson \& McCammon (1983) which assume neutral matter with standard abundances.
We found the results calculated from this code in good agreement with those of Magdziarz \& Zdziarski (1995).
As the mean free path of the X-ray photons in the disc is much shorter than any radius of curvature, the Euclidean metric approximation is generally applicable (George \& Fabian 1991), and we neglect GR effects on radiative transfer in the disc. Using the incident energy and direction as input, the photon is followed until it is absorbed or escapes from the disc.
When a photon escapes, the new constants of motion are computed from its
outgoing energy and direction.
Its trajectory toward the observer is solved and its energy and direction at infinity are
stored to build up the observed spectrum, unless the photon intercepts the disc again,
 in which case the Monte-Carlo reflection routine is used again,
 until it finally escapes to infinity
 or is absorbed in the disc.
This calculation method allows us to study
the effects of returning radiation and multiple-reflections that can not be considered by ray tracing methods.
These effects are therefore fully taken into account and will be discussed
on the basis of a few examples in Sect.~\ref{sec:returningrad}.

\begin{figure}[!h]
\centerline{\psfig{figure=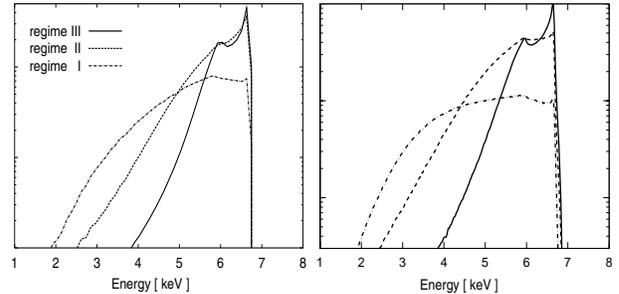,width=8cm,height=4cm}}
\centering
\caption{ The Iron $K_{\alpha}$ line profiles for an isotropic emitting disc in its rest frame calculated by a ray tracing method of Miniutti \& Fabian (2004) (left) and calculated by a direct method, from disc to observers at infinity, in GR for the ring-like primary source $\rho_s$ = 2 and $h$ = 2, 6 and 18 (right) that correspond to those of Miniutti \& Fabian (2004) in the regime I, II and III, respectively.}
\label{fig:compminiutti}
\end{figure}

We tested our gravitational shift calculation by computing and comparing the Iron $K_\alpha$ line profile with the results of Miniutti \& Fabian (2004), that are calculated by means of the ray tracing method.
We obtained a good agreement, see Fig.~\ref{fig:compminiutti}. In this simulation, the line emission was assumed to be locally isotropic to follow Miniutti \& Fabian (2004).
We stress however that our method can handle the (non-isotropic) angle dependence of the reflection component which, in the rest of the paper, is computed self consistently.

\section{Results for an on-axis Source}\label{sec:onaxis}

For the presentation of the results of the simulations we distinguish two different cases.
In this section we will consider the limiting case of a point source on
the rotation axis of the black hole (i.e. $\rho_{\rm s}$=0).
This simple situation often referred to as the "lampost model"
was considered by numerous authors in the context of disc illumination
and reflection/line studies (see Petrucci \& Henri 1997, Martocchia et al. 2000; Beckwith \& Done 2004, and references cited therein).
 It will enable us to illustrate the effects of the
height of the source on the shape and relative amplitude of the reflection continuum.
The more general case where the radius of the source ring is non zero will be treated in Sec.~\ref{sec:offaxis}
allowing us to study the effect of the radial distance of the source.
In the on-axis case we assume that the source is static.
The differences between the NN and
the SR are attributable only to the rotation of the disc.
The on-axis case is therefore simpler than the off-axis case, where the primary emission
is beamed due to the ring rotation, affecting strongly the angular distribution
of radiation observed at infinity and impinging on the disc
which in turn alters the reflected emission.
For both cases, we will compare the spectra obtained for the purely Newtonian,
Special Relativistic and fully general relativistic calculations,
considering successively the effects on the primary emission
(i.e. the observed luminosity originating directly from the hard X-ray source)
and the reflected component.

\subsection{Primary component}\label{sec:primary}

\begin{figure}[!h]
\centerline{\psfig{figure=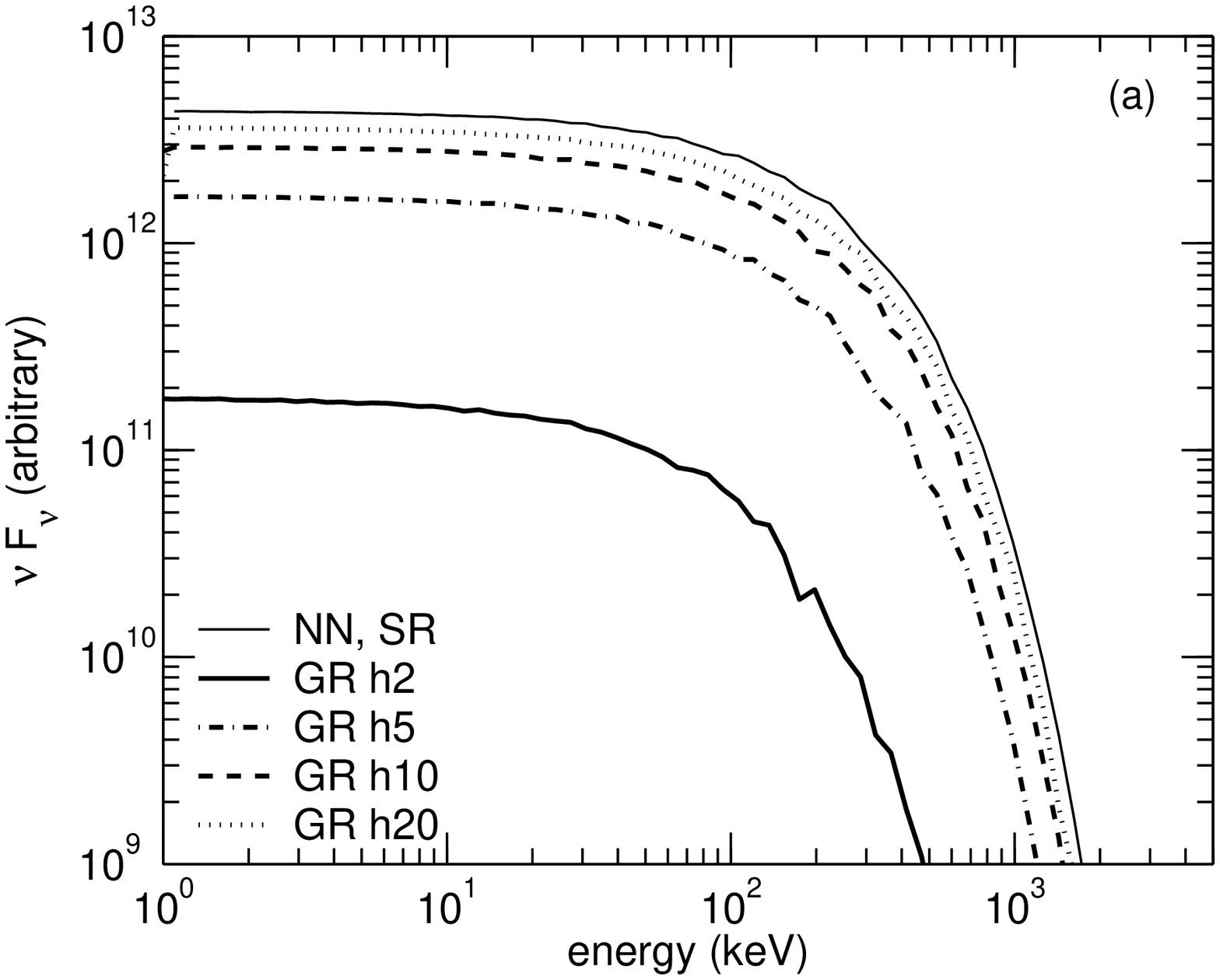,width=44mm,height=52mm}
\psfig{figure=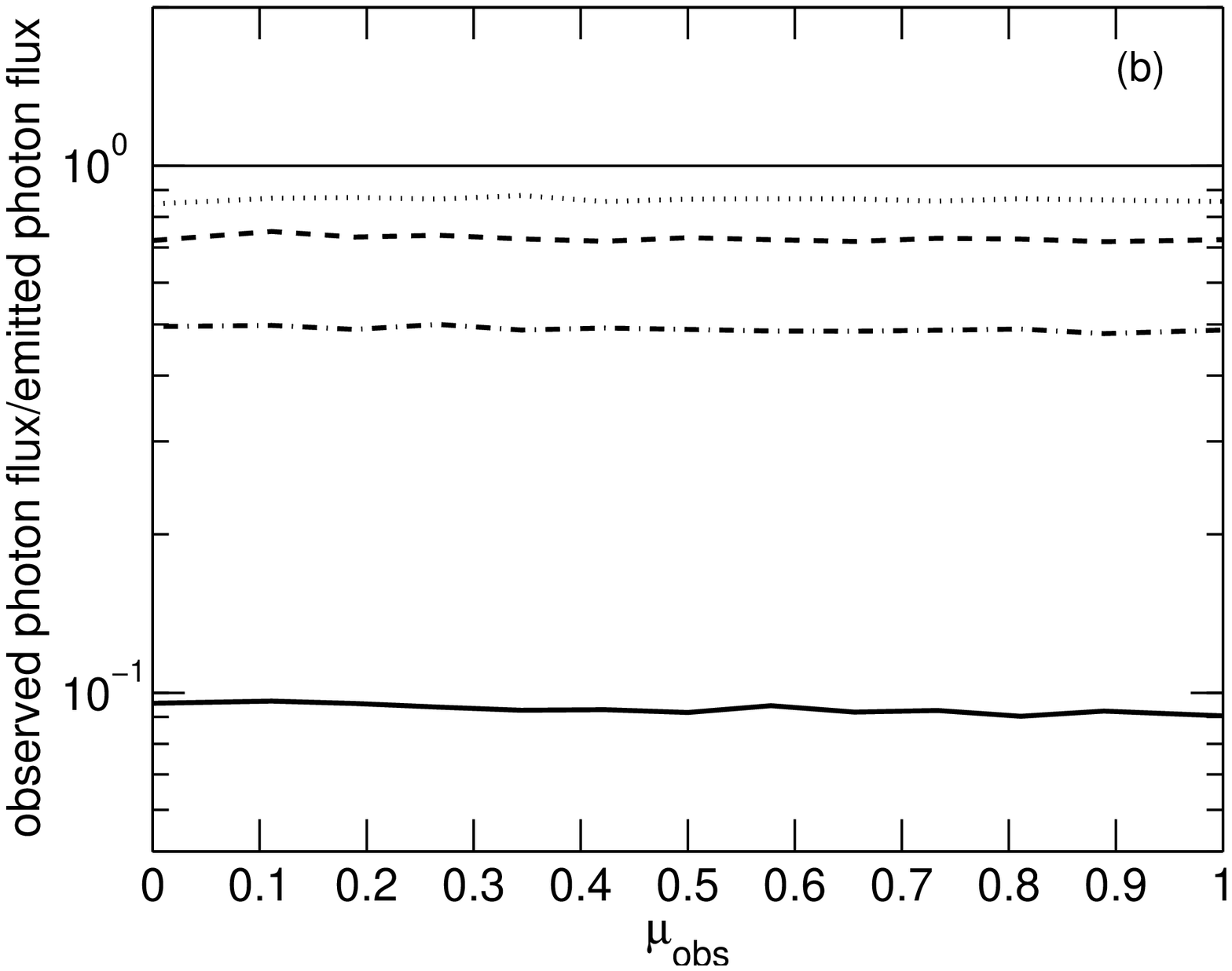,width=44mm,height=50mm}}
\caption{The primary spectra (a) and the primary total photon  flux as a function of
 $\mu_{obs}$ (b) of on-axis source models. The spectra and the primary total photon
 flux of GR with different source heights, 2, 5, 10 and 20, are shown
 from bottom to second line from top, respectively. The spectra and the primary
 total photon flux of NN and SR are equivalent and independent of the  source
  height as shown by the top line. }
\label{fig:onaxisprimary}
\end{figure}

Since the point source is static, the primary
spectra of NN and SR are identical as shown in Fig.~\ref{fig:onaxisprimary}.
The effects of space-time curvature manifest themselves through
spectral shift and the loss of primary component intensity.
For lower source heights, more photons are deflected toward the accretion
disc or to the horizon and the primary source appears fainter to an observer at infinity.
In addition, the gravitational redshift decreases the energy of the observed primary flux.
Obviously, these effects are more important for lower source heights.
In the GR models, the higher ($h=20$)
and the lower ($h=2$) source heights considered in Fig.~\ref{fig:onaxisprimary}
have an apparent luminosity differing by a factor of about 20.
The gravitational redshift for ($h=2$) is of order of $g=0.45$, as a consequence the cut-off energy is reduced
by a factor of about two with respect to the newtonian case.
Quite surprisingly,  although the photons travel across the rotating curved space-time,
the angular distribution of the primary component at infinity is very close to isotropic
whatever $h$, as shown in the left panel of Fig.~\ref{fig:onaxisprimary}.

\subsection{Reflection component}\label{sec:reflect}

\begin{figure}[!h]
\centerline{\psfig{figure=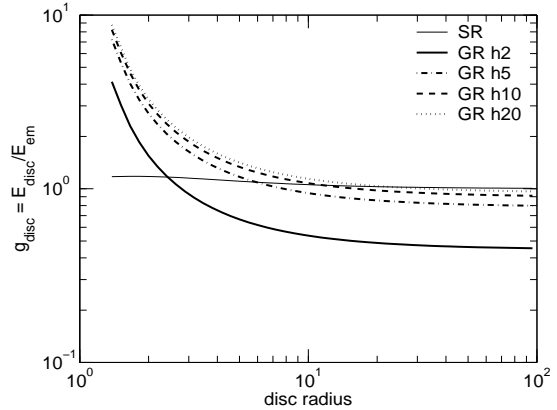,width=73mm,height=55mm}}
\centering
\caption{Ratio of the energy locally measured in the disc frame to the emitted energy from the primary source, $g_{disc}$, as a function of disc radius with different source heights, 2, 5, 10 and 20, (in the outer region) from bottom to second line from top, respectively. This ratio in the SR case, thin solid line, is independent of the source height. }
\label{fig:gdisc}
\end{figure}

Since the shape of the reflected component depends on the spectrum and angular distribution
of the impinging radiation, it is interesting to study the effects
of special and general relativity on the irradiation as seen in the disc rest frame.
Fig.~\ref{fig:gdisc} shows the gravitational photon energy shift between  the source and disc frame, $g_{disc}$=$E_{disc}/E_{em}$ as a function of disc radius.
In the inner disc region (approximatively for $r<h$) the impinging photons
have been falling deeper into the potential wheel and have therefore gained energy.
In this region the material in the disc sees a harder spectrum.
The amplification factor can be of almost 10 at the inner stable orbit.
This gravitational effect is much stronger than the simple Doppler boosting
due to disc rotation which leads at most to a 20 \% gain in energy as can be seen
on Fig.~\ref{fig:gdisc}.
On the other hand, in the outer part of the disc the impinging photons have
travelled against the gravitational potential and the radiation is redshifted.

\begin{figure}
\centerline{\psfig{figure=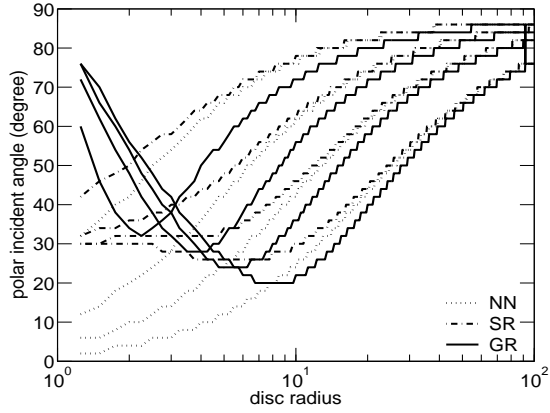,width=73mm,height=55mm}}
\centering
\caption{Photon incident angle of three models as a function of disc radius
with different source heights, 2, 5, 10 and 20, from left to right, respectively.
Normal incidence is $0^o$, grazing incidence is $90^o$.}
\label{fig:angledisc}
\end{figure}

The disc rotation and the rotating curved space-time have also an effect on the photon
incident angle in the inner region.
Fig.~\ref{fig:angledisc} compares the incident angle of the photons as a function of radius obtained for the NN,
SR and GR cases.
In the newtonian static case, the incident angle $\theta_{\rm i}$
(with respect to the disc normal) is simply given by $\theta_{\rm i}=\arctan{r/h}$.
The aberration of light caused by the disc rotation in SR leads to a higher inclination than
in the static disc NN.
The effect is stronger in the inner part of the disc where the rotation velocity is larger.
In the Kerr metric, light bending causes the incidence angle to be smaller than in the newtonian case at large radii.
On the other hand, the photons emitted toward the inner parts of the disc ($r<h$) are dragged by the rotation of the metric and start
spiraling fast around the axis of the black hole. Those photons impinge the innermost parts of the disc with an almost grazing incidence,
so that at short radii the incident inclination of photons is higher than both in the SR and NN.

\begin{figure}[!h]
\centerline{\psfig{figure=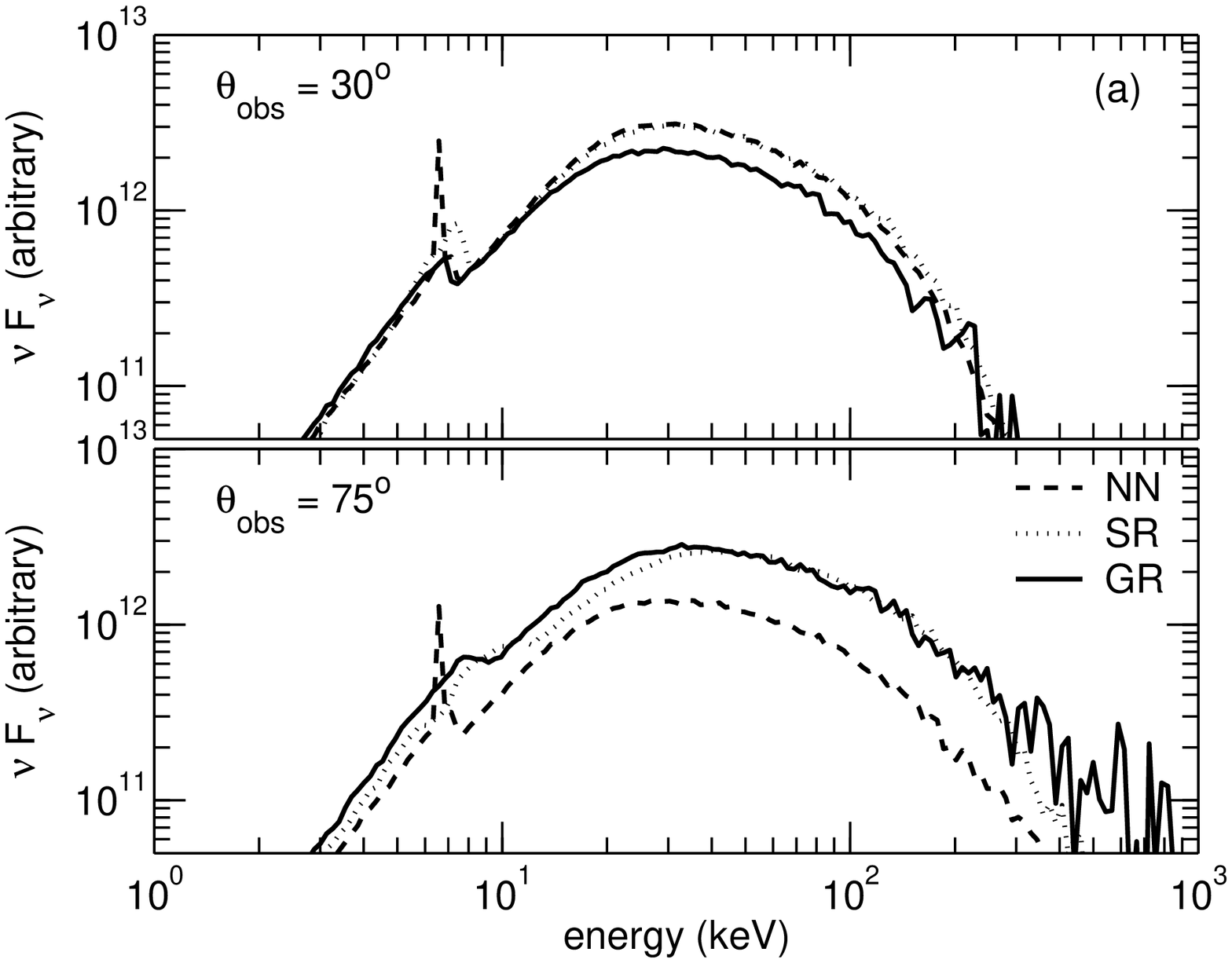,width=44mm,height=50mm}\psfig{figure=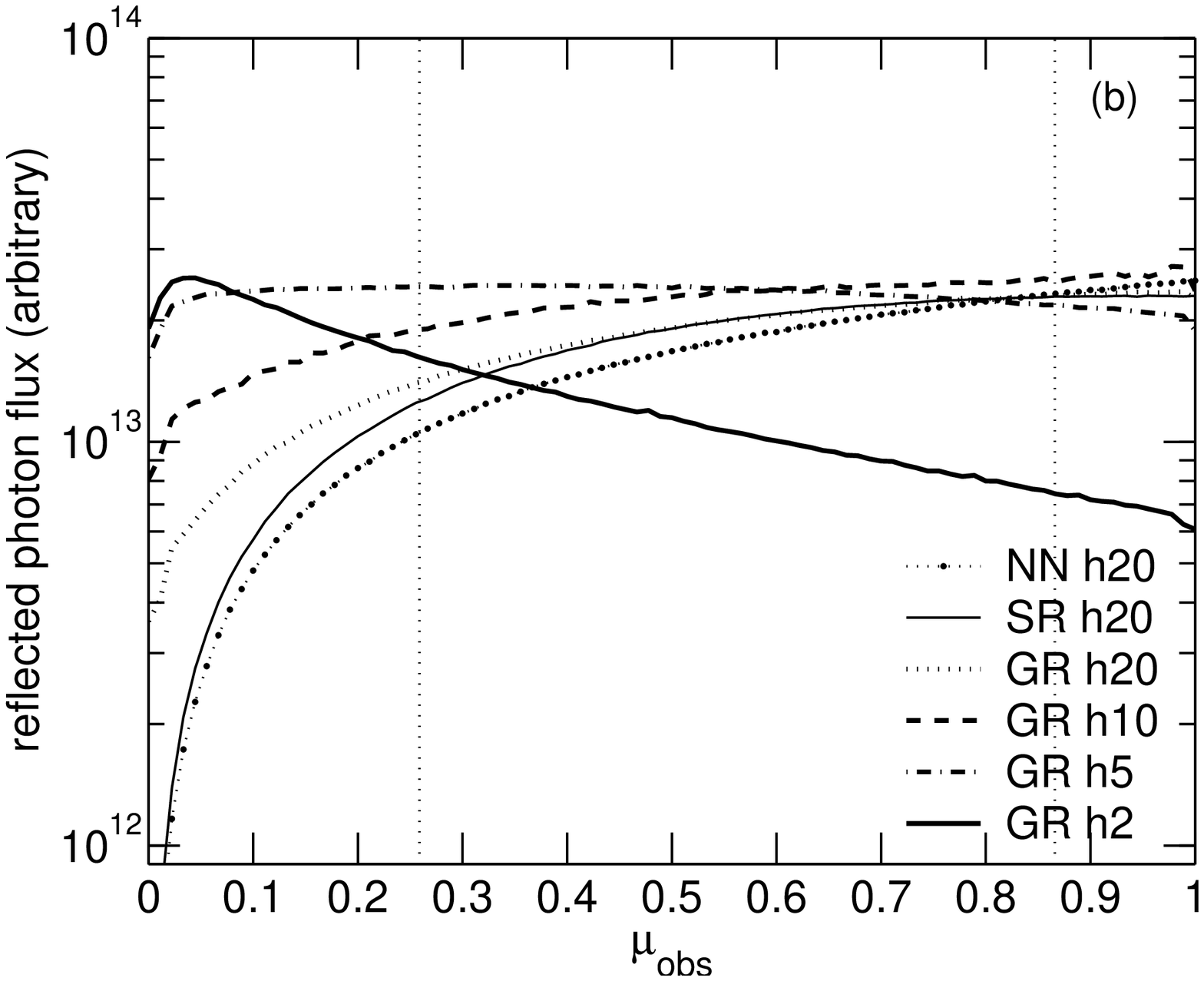,width=44mm,height=50mm}}
\centering
\caption{(a) Predicted reflected spectra  of NN (dashed line), SR (dashed-dotted line)
and GR (solid line) for $\theta_{obs}$ = $30^o$ (top) and $\theta_{obs}$ = $75^o$
(bottom) for $h=5$.
(b) The reflected photon flux as a function of $\mu_{obs}$ For GR with different source heights, 2, 5, 10 and 20, from top to third line from bottom, respectively. For $h$ = 20, we show those of NN and SR, bottom line and the next for high inclinations, respectively.}
\label{fig:refonaxis}
\end{figure}

 Fig.~\ref{fig:refonaxis}(a) shows examples of reflection spectra computed
according to our three fiducial models and two different inclinations of the line of sight.
At low inclination, the SR effects are weak and the NN and SR spectra are very similar.
The GR spectrum has a lower amplitude and lacks high energy photons due to gravitational
redshift. On the other hand, for large inclinations, the
spectral shifts are more important. The iron line in the SR spectrum  is significantly blurred, and the overall spectrum is blueshifted.
The GR spectrum shape is then similar to the SR spectrum, the Doppler effects
dominate over gravitational redshift.
Their normalisations however differ due to the effects of light bending.

We notice also that the reflected spectra of GR have more photons at high energy for high inclination, as shown in Fig.~\ref{fig:refonaxis}(a). Indeed, the incident photon energies, in the disc frame, in the central area are amplified by the gravitational shift. Most of these reflected photons are observed at large inclinations. By means of the reflected spectra of NN as a reference, the spectra of GR shift to the red side for low inclinations and to the blue side for high inclinations. The importance of spectral shifts increases with the lowness of the source location.

Fig.~\ref{fig:refonaxis}(b) shows the angular distribution of the reflected energy
flux for different models.
The intrinsic angle dependence of the Compton reflected component already
differs significantly from isotropy with more photons
reflected toward highly inclined lines of sight.
In the SR case, the beaming effects due to the rotation of the disc lead
to an overall larger reflected luminosity. Those beaming effects also affect
the shape of the angular distribution by
enhancing the reflected flux at higher inclination (i.e. in the directions
close to the direction of motion of the material in the disc).
This trend (larger reflection component at larger inclination) is further enhanced by the
GR effects. The effects of light bending and gravitational shift are two-fold:
First, the number and average energy of the illuminating photons are larger, as a result the total reflected flux
is larger. Second, the path of the reflected photons is deflected toward higher inclinations.

Contrary to the NN and SR cases,  in the GR model
the angular distribution of the reflected component strongly depends
on the source height. For lower source heights, the stronger
illumination increases the reflected flux (by a factor of about two between $h=20$
and $h=5$). However when the  X-ray source is very close to the black hole
most of the radiation is lost  in the hole and the reflected
luminosity is decreased (by almost 40 $\%$ between $h=5$
and $h=2$).  Because the illumination of the central parts of the disc is stronger
at low source heights,  the light  bending of reflected photons
toward large inclination angles is then dramatically
enhanced. So that, for $h=5$ or lower,
the angular distribution peaks above $75^o$, as shown in Fig.~\ref{fig:refonaxis}(b). This behaviour
is qualitatively very different from the Newtonian case where
the observed flux is stronger at low inclination.

\subsection{Both components}\label{sec:onaxisboth}

\begin{figure}[!h]
\centerline{\psfig{figure=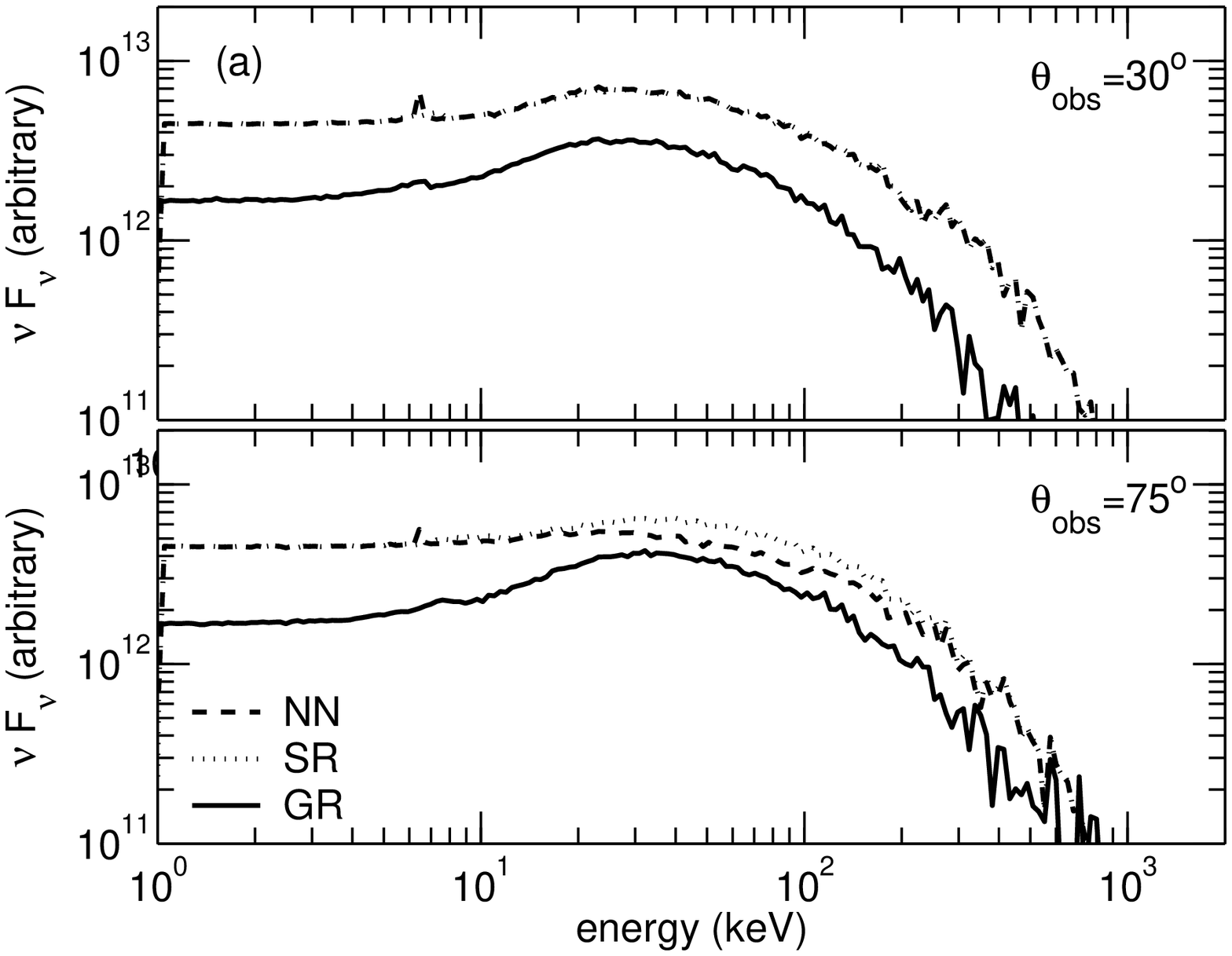,width=44mm,height=48mm} \psfig{figure=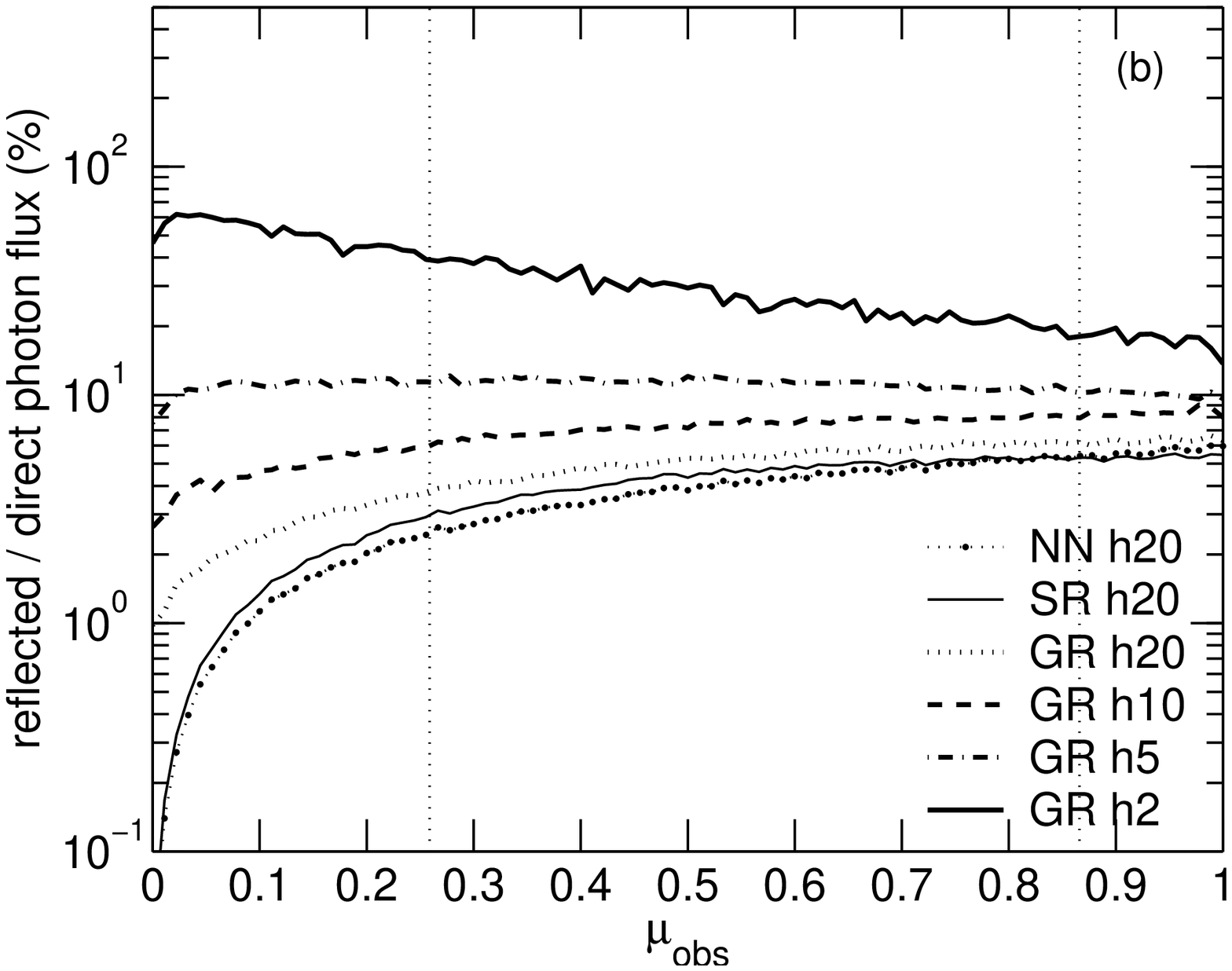,width=44mm,height=48mm}}
\centering
\caption{(a) Predicted spectra, shown as the same of Fig.~\ref{fig:refonaxis}(a). (b) The fraction of reflected total photon flux to primary total photon flux expressed as a percentage as a function of $\mu_{obs}$ , shown as the same of Fig.~\ref{fig:refonaxis}(b). }
\label{fig:onaxistotal}
\end{figure}

In this section, we combine the results from the Sec.~\ref{sec:primary}
and \ref{sec:reflect} to study the dependence of the total (primary+reflected) spectrum
on the parameters of the model.
Fig.~\ref{fig:onaxistotal}(a) shows the total spectra obtained for our
 three fiducial models at inclinations of $30^o$ and $75^o$.
 Since the NN and SR spectra are dominated by the primary component they
 depend only weakly on the inclination angle.
 We can note however that the reflected features
 appear weaker at large inclination.
 The difference between SR and NN models appears only at high inclination
 when the reflection component is more strongly smeared.

 On the other hand
 the GR spectra have a reduced normalisation due to deflection of photons towards the disc
  and gravitational redshift.
  Moreover contrary to the NN and SR models
  the reflection bump appears much stronger at high inclination.
  This is a consequence of the light bending effects on the reflection component (see
  Sec.~\ref{sec:reflect}), while, as mentioned in Sec.~\ref{sec:primary}
 the primary emission remains almost isotropic.

 Let us consider further the angular dependence of the relative strength of the reflection.
 Fig.~\ref{fig:onaxistotal}(b) shows the angular distribution of the fraction
 of the reflected flux to the primary flux,
 hereafter RPF. This figure shows that in all GR models the RPF is higher than in the simple
 NN or SR models. The relative amplitude of the reflection component strongly increases
 for lower source heights, specially at large inclinations, as discussed above.

\begin{figure}[!h]
\centerline{\psfig{figure=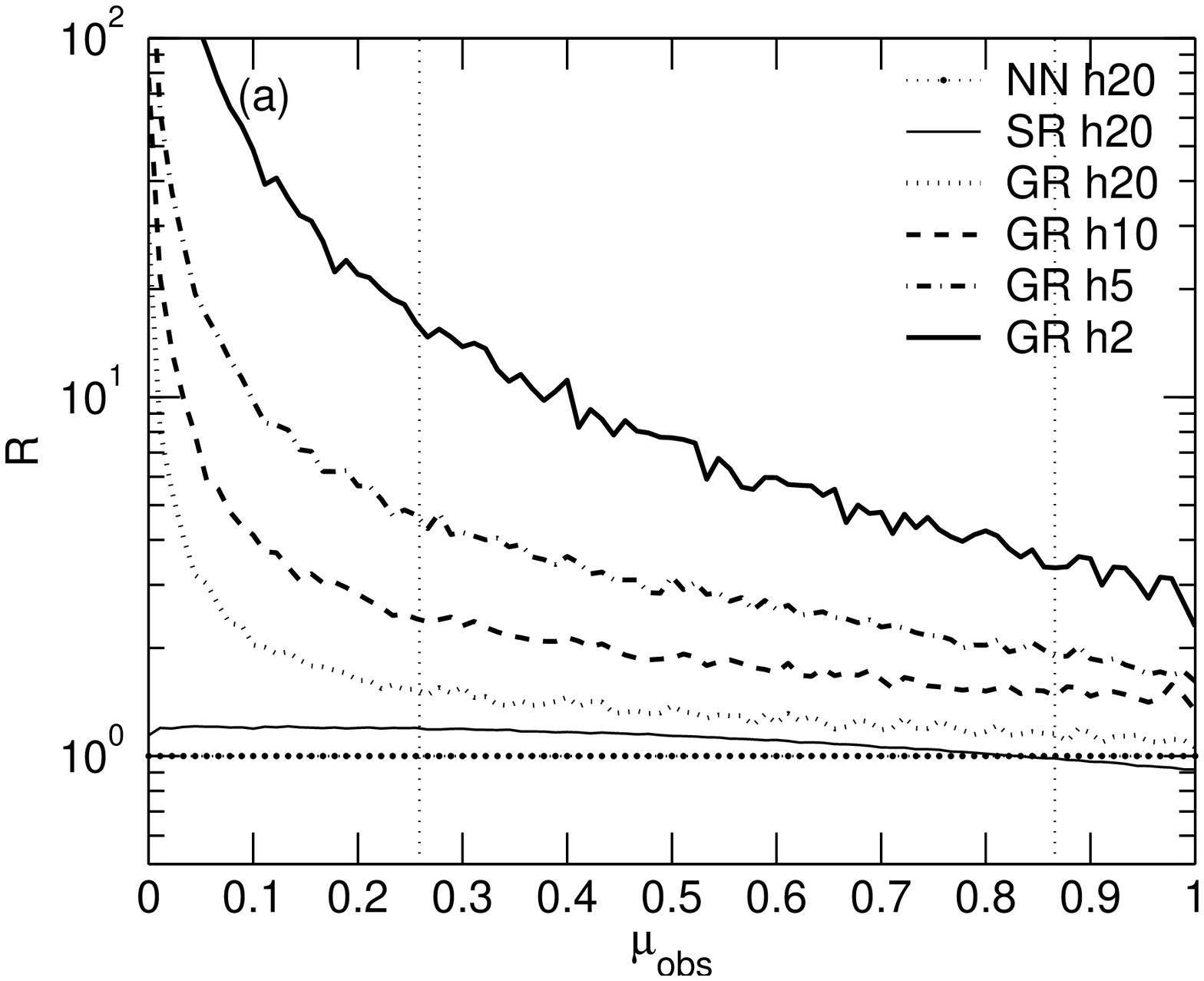,width=44mm,height=46mm}
\psfig{figure=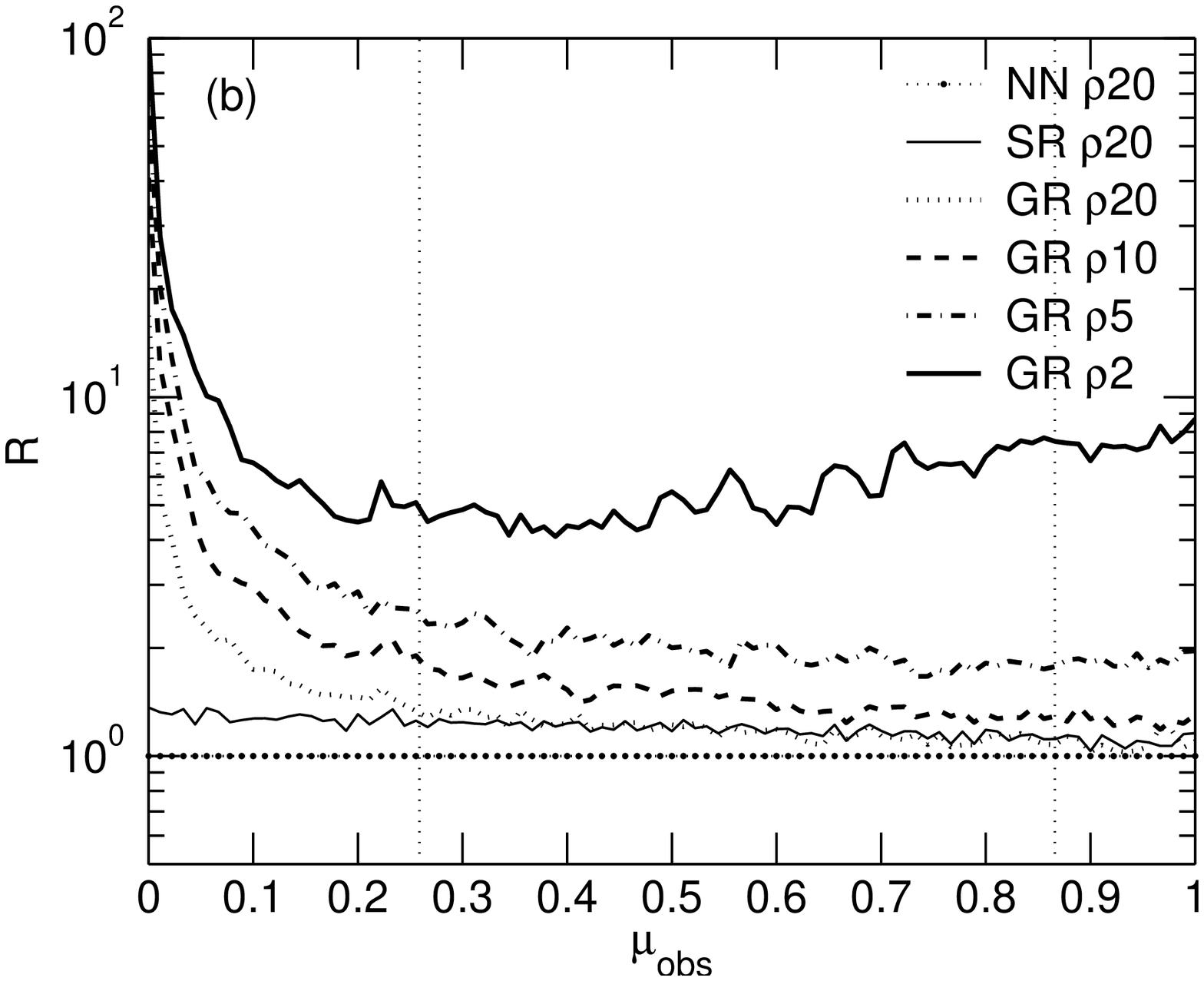,width=44mm,height=46mm}}
\centering
\caption{$R$ as a function of inclination, (a) for the on-axis model with different source heights, (b) for off-axis model with different source radii for a fixed source height at 2.}
\label{fig:rdemu}
\end{figure}

 For the purpose of the comparison with data it is useful to transpose our results
  in terms of the usual reflection coefficient $R$
  commonly used to quantify the relative fraction of reflection in observed spectra.
  By definition $R$ is unity for an isotropic source above an infinite reflector.
  This situation is closely approached for our NN model at large source height.
  Therefore we can estimate $R$ for our different models
  by simply computing the ratio of their RPF  to that of the NN model.
  The result is shown in Fig.~\ref{fig:rdemu}.
  In GR models $R$ is always larger than unity. It increases with
  both higher inclinations and lower source heights up to
  huge values (10-1000) in the most extreme cases.

\section{Results for a ring-like source}\label{sec:offaxis}
In this section, we study the effects of the co-rotating source in flat space-time and
 in rotating curved space time with different heights, $h$, and different radii, $\rho_s$.
\subsection{Primary component}\label{sec:primary2}

\begin{figure}[h]
\centerline{\psfig{figure=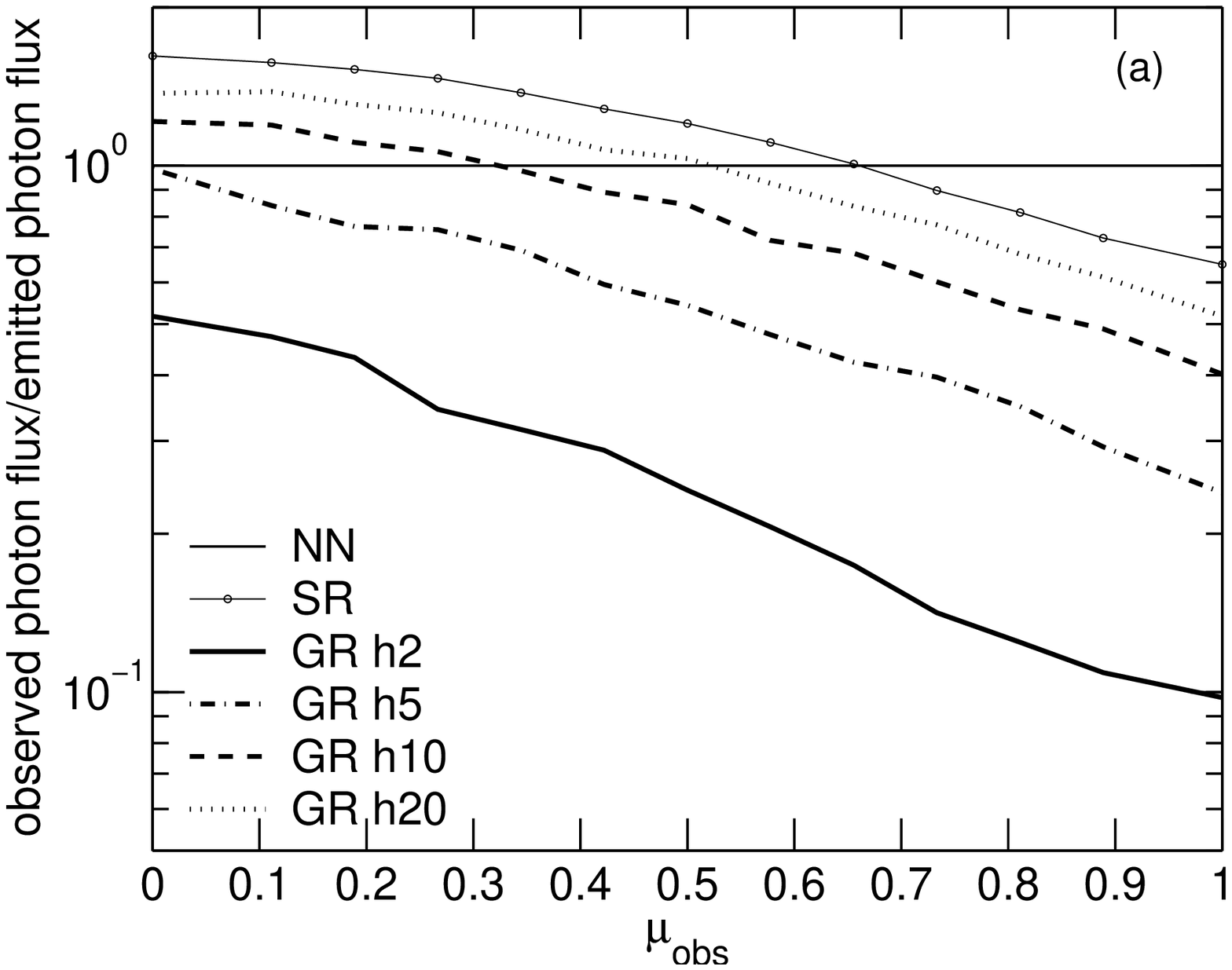,width=44mm, height=46mm}\psfig{figure=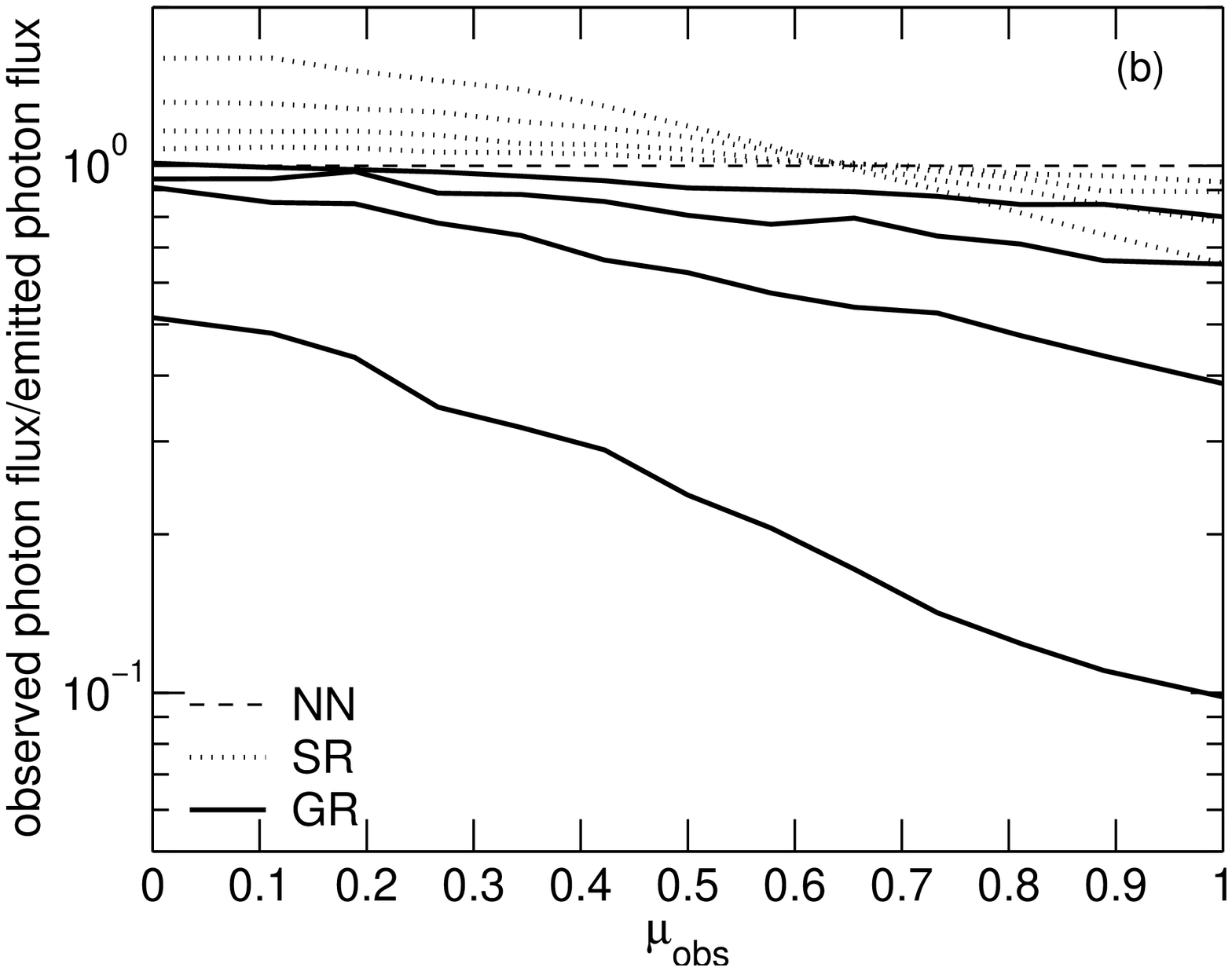,width=44mm, height=50mm}}
\centering
\caption{(a) The primary total photon flux as a function of $\mu_{obs}$
with fixed $\rho_s$ = 2  and different source heights, 2, 5, 10 and 20, are shown
from bottom to second line from top, respectively.
 The spectra and the primary component of SR are independent of the source height
 as shown by the top line. For NN, the observed primary component
 is isotropic.
  (b) The same as (a) but with fixed source height at 2 and different $\rho_s$, 2, 5, 10 and 20 from bottom to top, respectively.}
\label{fig:offaxisprimary}
\end{figure}

Contrary to the static on-axis case, in the off-axis source model, the rotation of the
ring source affects the primary component through Doppler and beaming effects.
In particular significant changes are expected in the angular distribution
of the observed primary radiation, as can be seen in Fig.~\ref{fig:offaxisprimary}.
The SR calculation shows that the effect of rotation is to beam the emission
 toward large inclination angles, as expected.
 This effect is stronger at lower distances to the axis $\rho_{s}$
 i.e. at larger disc/ring velocities (see Fig.~\ref{fig:offaxisprimary}(b)).
 We obtain almost identical results for the GR model except that the total number
  of photons reaching infinity is lower due to light bending.
   As in the on-axis model, in GR, the reduction
  of the primary component is stronger for lower source heights,
  but contrary to the on-axis model
  the observed angular distribution of radiation is no longer isotropic (see Fig.~\ref{fig:offaxisprimary}(a)).

\subsection{Reflection component}

\begin{figure}[h]
\centerline{\psfig{figure=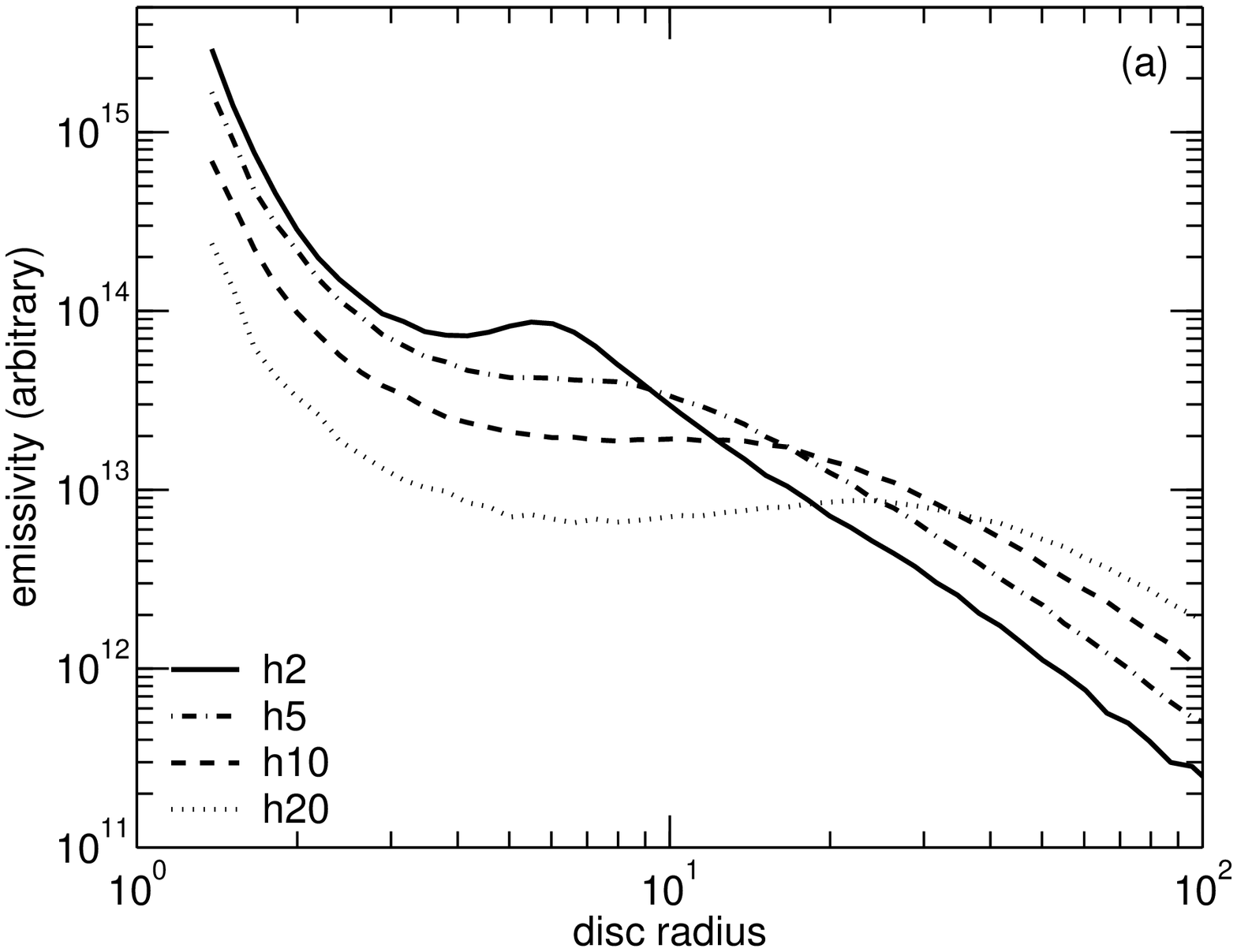,width=44mm, height=48mm}\psfig{figure=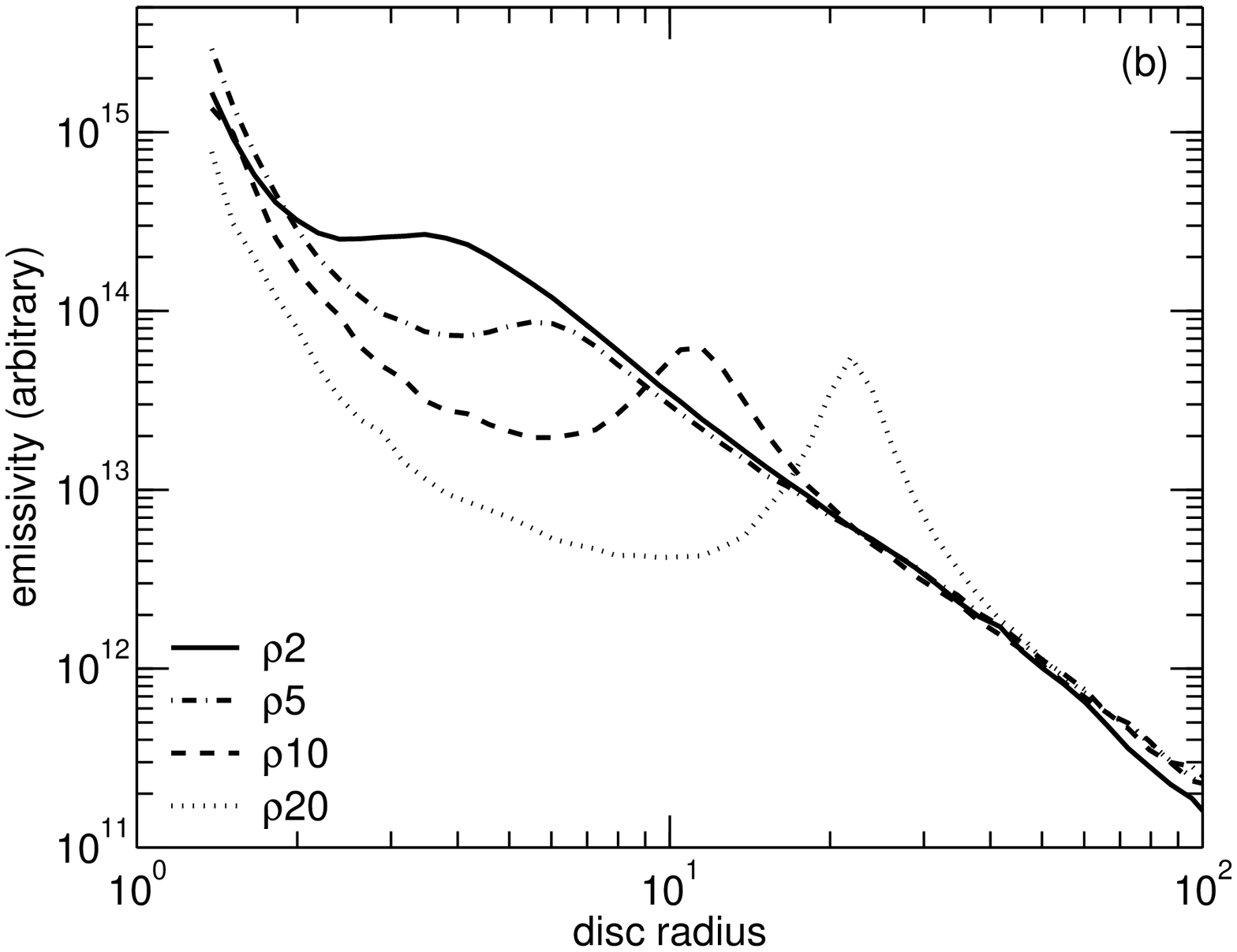,width=44mm, height=48mm}}
\centering
\caption{The reflected emission (disc emissivity) of GR as function of disc radius for a fixed $\rho_s$ = 5 with different $h$ (a) and for a fixed $h$ = 2 with different $\rho_s$ }
\label{fig:emisoffaxis}
\end{figure}

\begin{figure}[!h]
\centerline{\psfig{figure=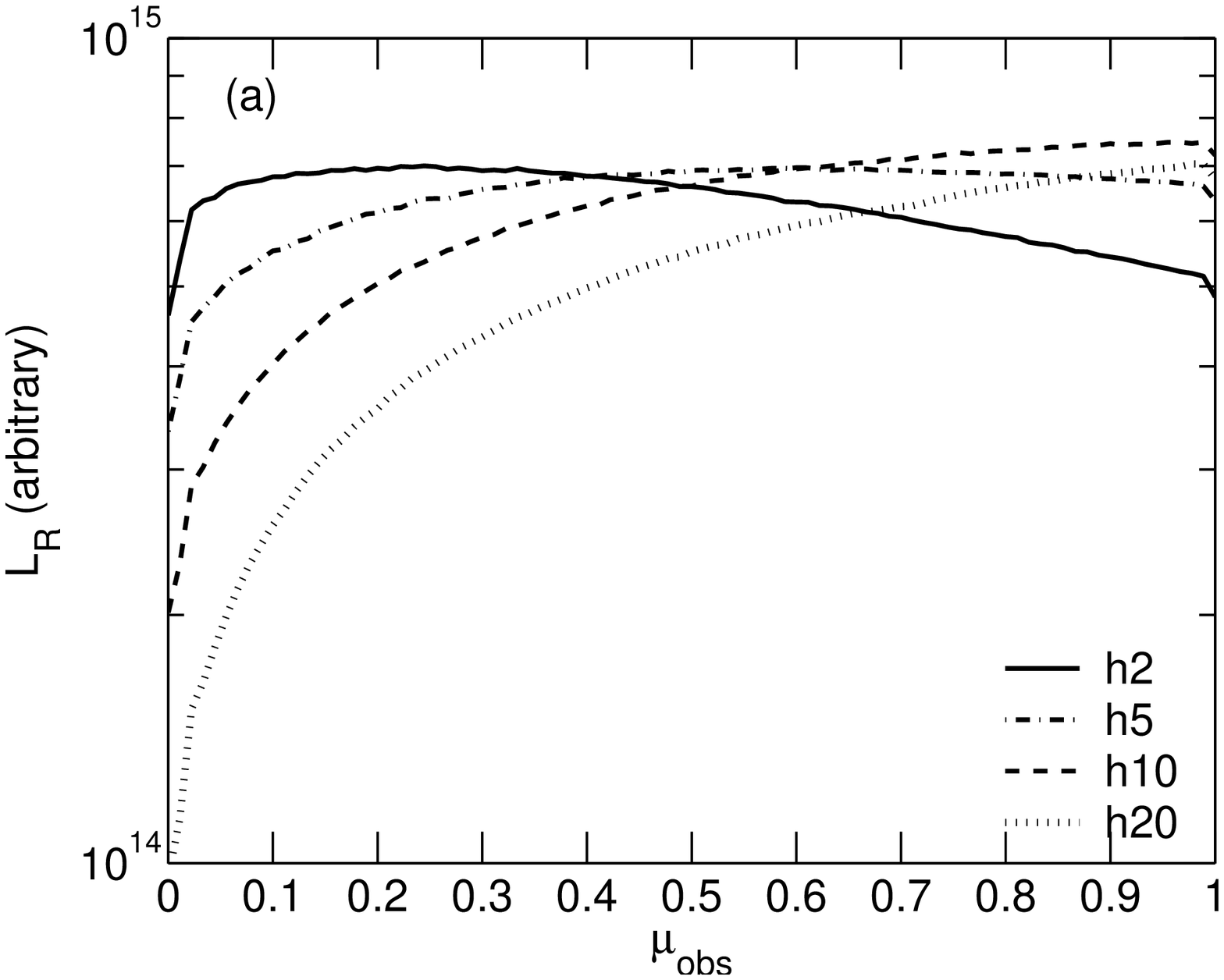,width=44mm, height=48mm}\psfig{figure=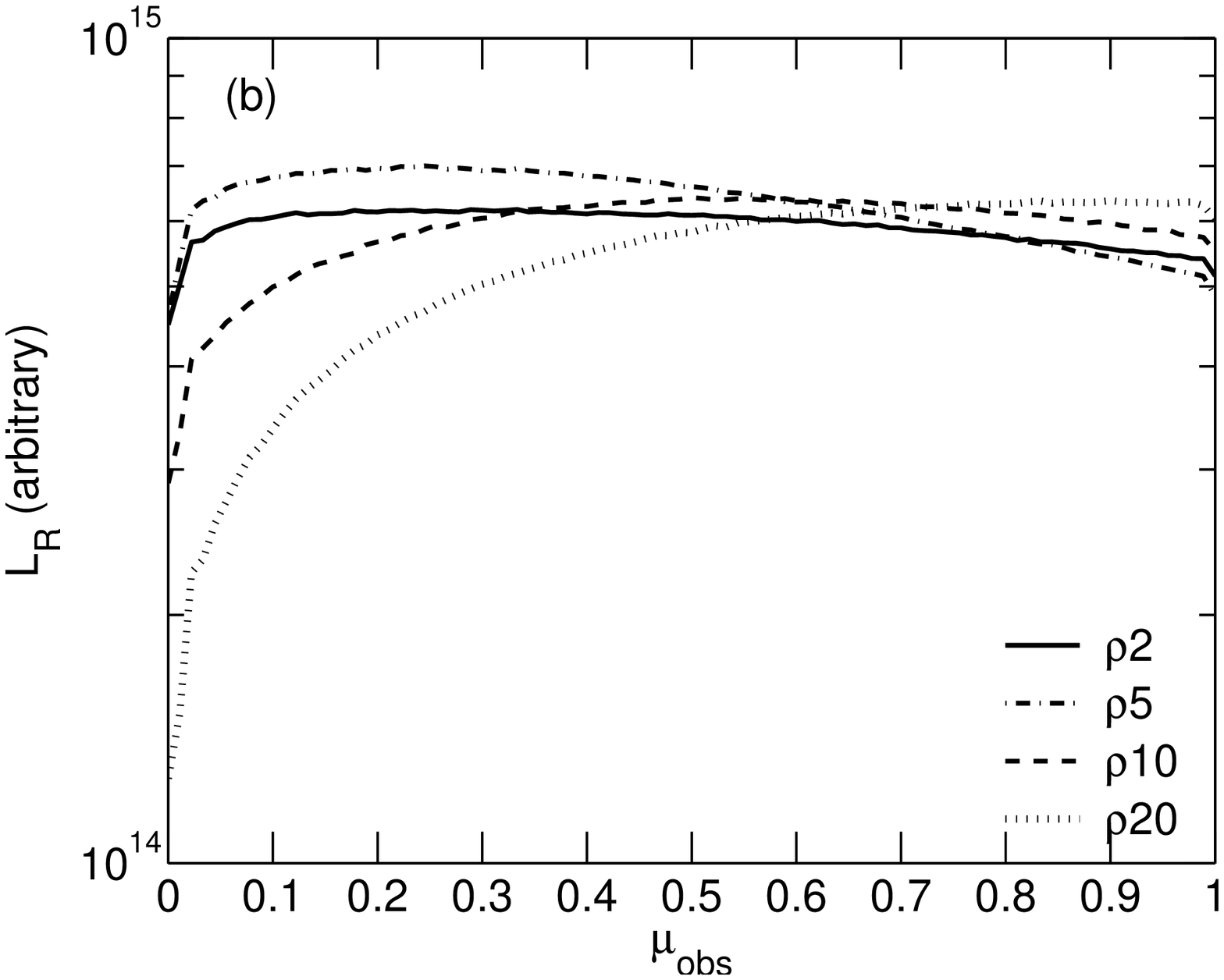,width=44mm, height=48mm}}
\centering
\caption{The reflected total energy flux as a function of $\mu_{obs}$ for a fixed $\rho_s$ = 5 with different $h$ (a) and for a fixed $h$ = 2 with different $\rho_s$ }
\label{fig:refoffaxis}
\end{figure}

What is the effect of varying ring inner radius on the reflected component?
 Fig.~\ref{fig:emisoffaxis} shows the disc emissivity for various values
   of the radius and height of the ring.
   For shorter $\rho_{\rm s}$ (and/or $h$) the
   emission of the source is more sensitive to GR effects that tend
  to deflect its emission toward the disc.
  As a consequence, the overall illumination and reflected flux are larger.
   In addition,  at short $\rho_{\rm s}$ (and/or $h$)
    the emissivity of the central parts of the disc is larger.
  This means that most of the reflection is produced close to the black hole,
   leading to strong light bending toward the highly inclined lines of sight.
  Fig.~\ref{fig:refoffaxis} shows the angular distribution of
the reflected radiation in the off-axis model for different source heights and ring radii.
As discussed above, the qualitative dependences of the angular distribution of the reflected
 radiation on the ring radius   $\rho_{\rm s}$ and height of the source $h$
 are very similar: smaller values lead to larger reflection fluxes strongly
 deflected toward large inclination observers.

\subsection{Both components}

\begin{figure}[!h]
\centerline{\psfig{figure=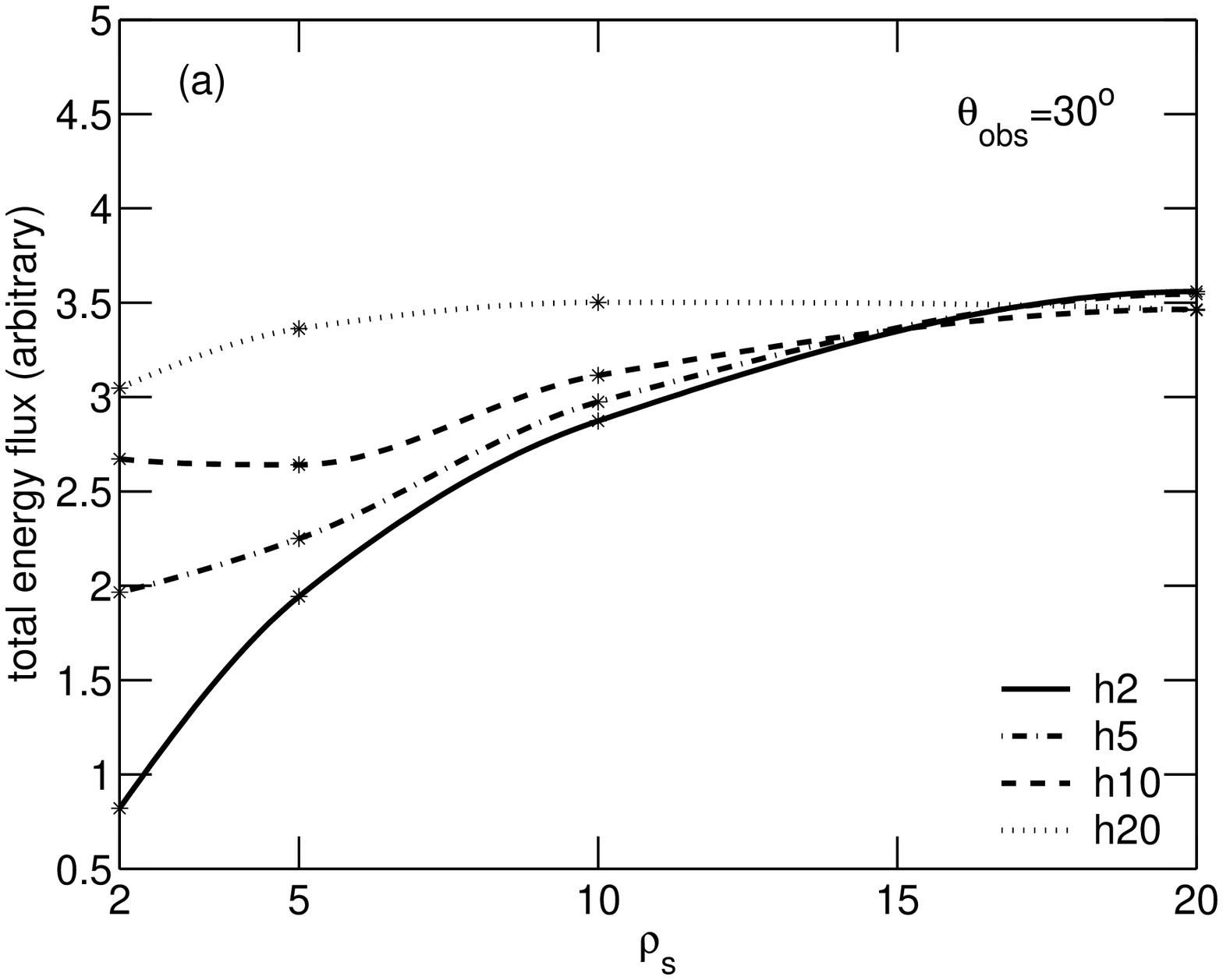,width=44mm, height=48mm}\psfig{figure=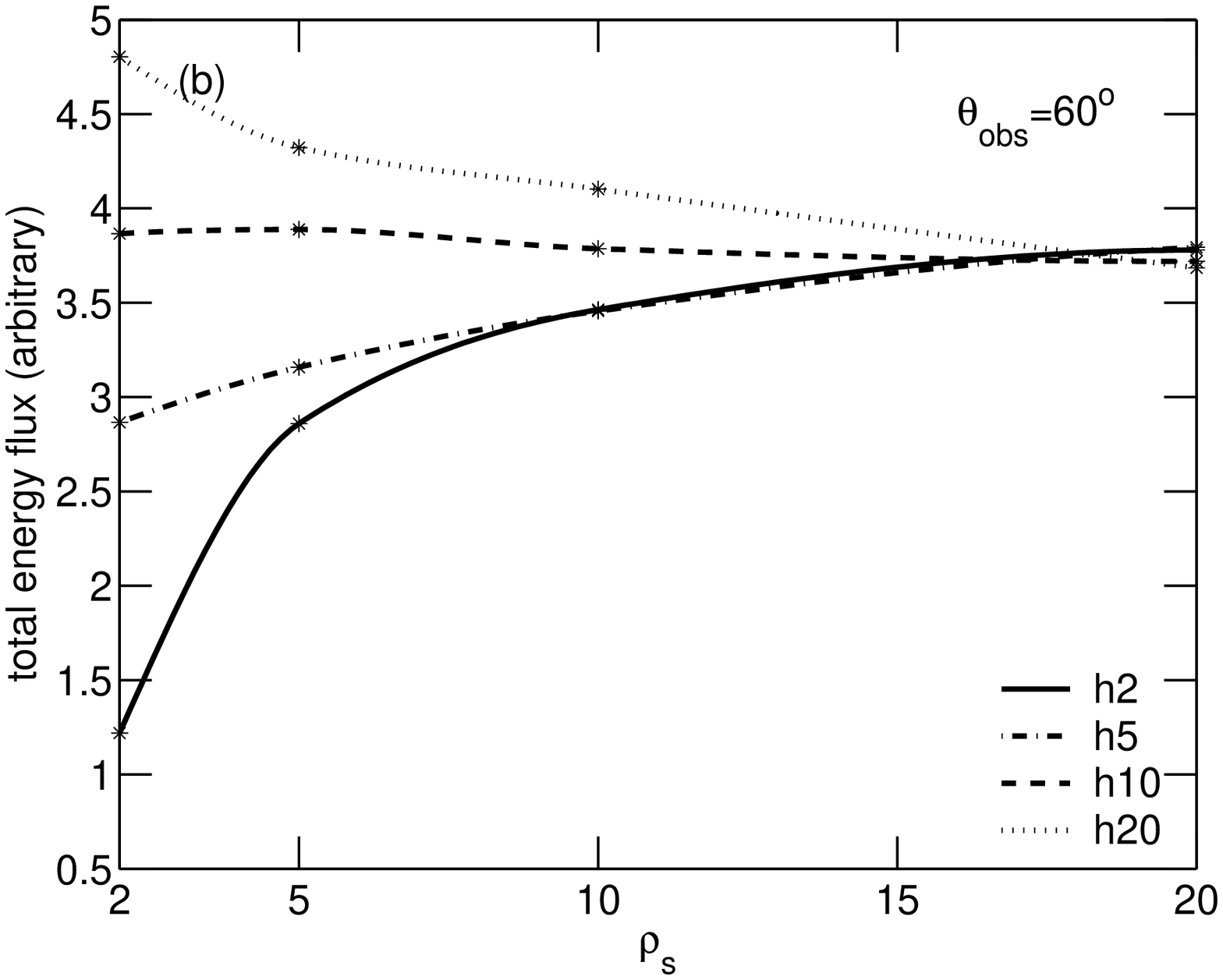,width=44mm, height=48mm}}
\centering
\caption{The predicted total energy flux, which of primary components plus reflected components, in energy range 1-30 keV with different $h$, presented by different lines, as function of $\rho_s$ for an inclination of $30^o$ (a) and $60^o$ (b) [see text for details].}
\label{fig:varibility}
\end{figure}

\begin{figure}[!h]
\centerline{\psfig{figure=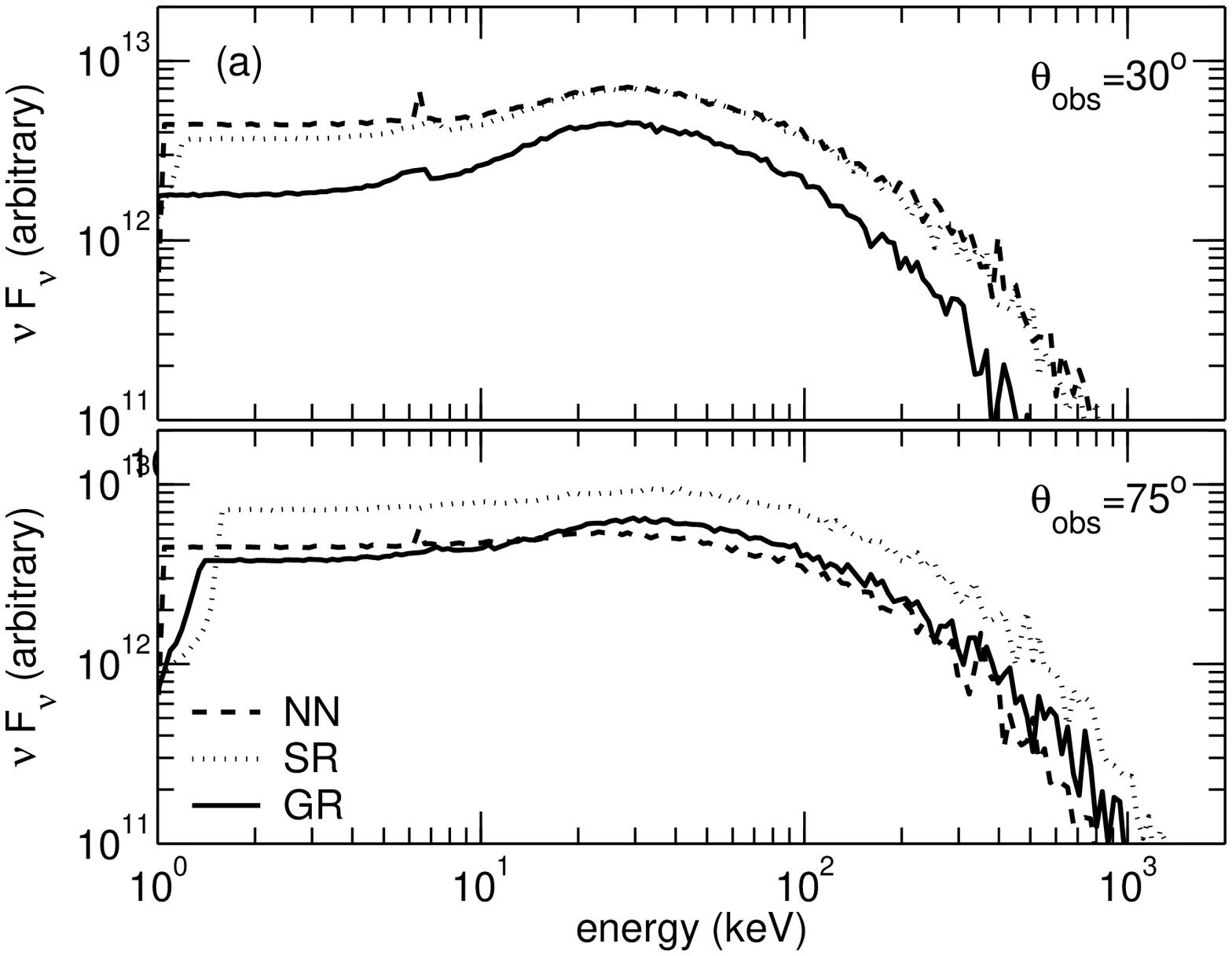,width=44mm, height=48mm}\psfig{figure=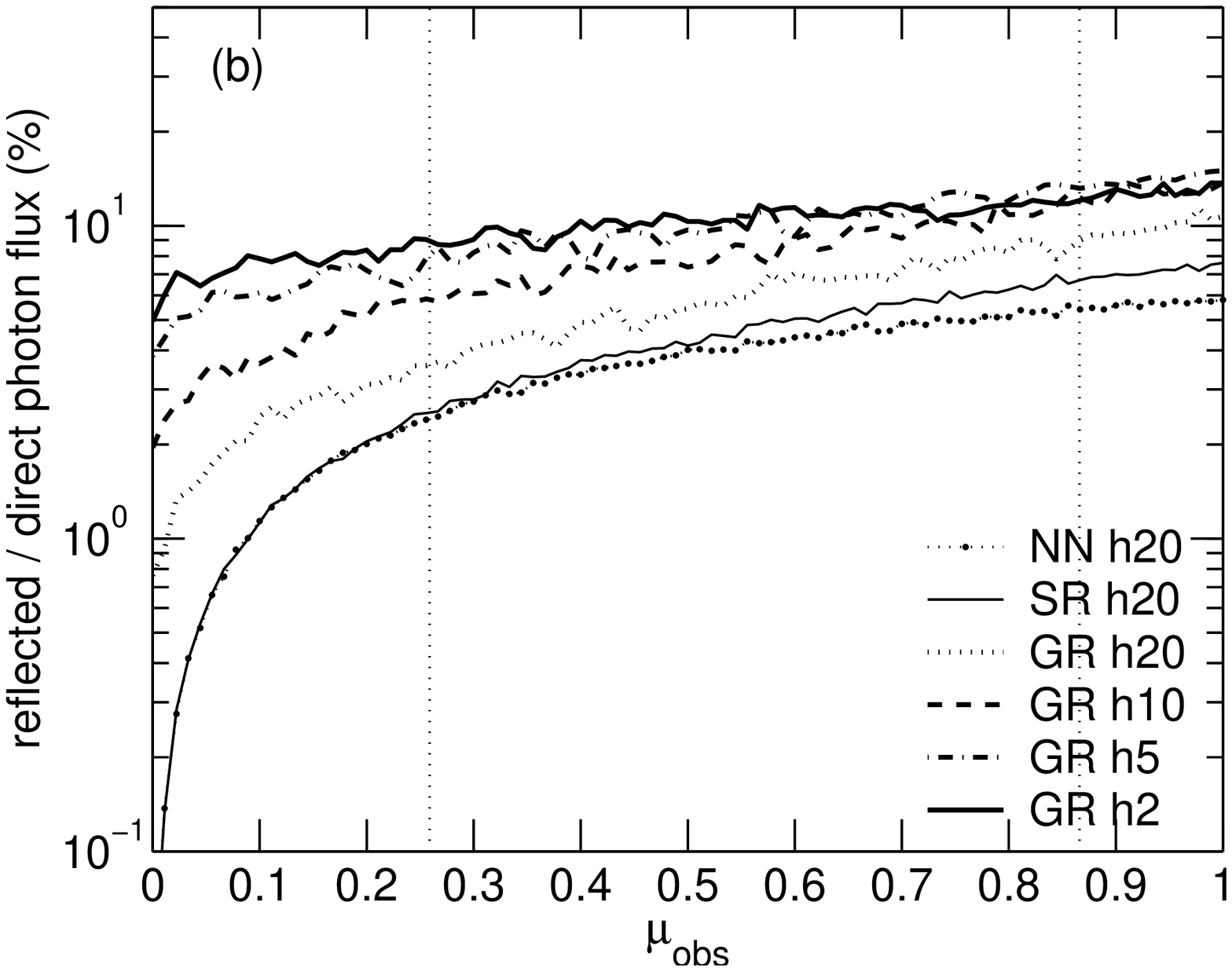,width=44mm, height=48mm}}
\centering
\caption{Shown as the same of Fig.6 but for the off axis source cases with $\rho_s$ = 5 and $h=5$. The artificial cut-off in energy range 1-2 keV results from our minimum energy in the rest frame of the primary photons under the effects of the gravitational blue-shift. }
\label{fig:offaxistotal}
\end{figure}

The dependence of total observed luminosity, in energy range 1-30 keV, on the ring source radius are shown in Fig.~\ref{fig:varibility}.
As can be seen on this figure, the variations of the observed flux with $\rho_{s}$ are strongly angle dependent.
For an inclination of $30^o$ at $h$ = 2, the luminosity drops by about 45\% when the ring source radius decreases from 20 to 5 and
only about 25\% for $60^o$.
For $h=20$, the luminosity even increases when $\rho_s$ is reduced from
20 to 2 for $60^o$.
There are similar effects when the source height diminishes for the fixed radius source.
Fig.~\ref{fig:offaxistotal}(a) compares total spectra obtained for the NN, SR, and GR models.
The main difference with the results obtained for the on-axis model (Fig.~\ref{fig:onaxistotal})
is that, at high inclinations the observed primary flux is boosted by the rotation of the ring
(as discussed in Sec.~\ref{sec:primary2}). This beaming effect  compensates the drop
 in luminosity due to light bending and at $75^o$ the normalisation of the NN
 and GR models are comparable.
 This strong Doppler beaming of the primary emission leads
  to a reduction of the RPF at large inclination at all $h$ and $\rho_{\rm s}$ as shown in Fig~\ref{fig:offaxistotal}(b).
 This is an important qualitative difference with the on-axis model where the overall
  spectra can become increasingly reflection dominated at large inclination.
  However the decline in the relative fraction of the reflected flux at large inclinations
   is less pronounced than in the NN model, and therefore
  the reflection coefficient $R$ (defined as the GR to NN RPF ratio,
  see Sec.~\ref{sec:onaxisboth}) can still increase at large inclinations
  (see Fig.~\ref{fig:rdemu}(b)).


\subsection{Effects of returning radiation}\label{sec:returningrad}

\begin{figure}[h]
\centerline{\psfig{figure=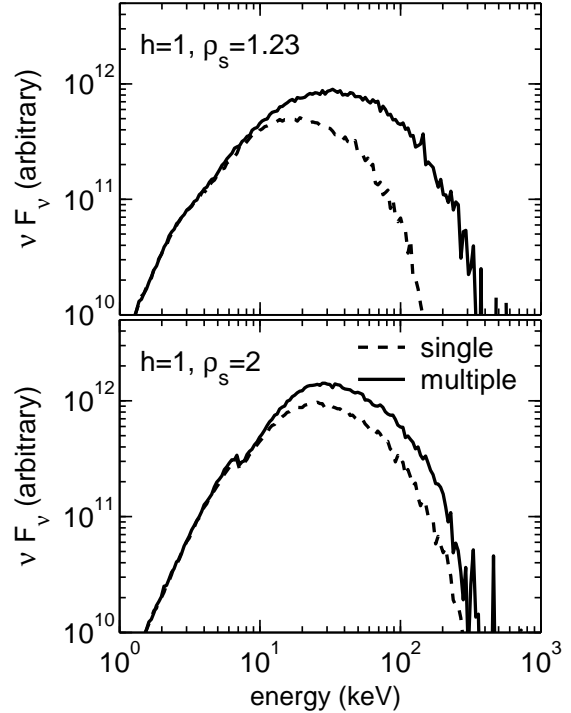,width=75mm}}
\centering
\caption{Effects of returning radiation on the reflected spectrum
for the inclination of $30^o$, solid curves: total multiple reflection spectrum, dashed curves:
single reflection only.}
\label{fig:returnrad1}
\end{figure}

\begin{figure}[h]
\centerline{\psfig{figure=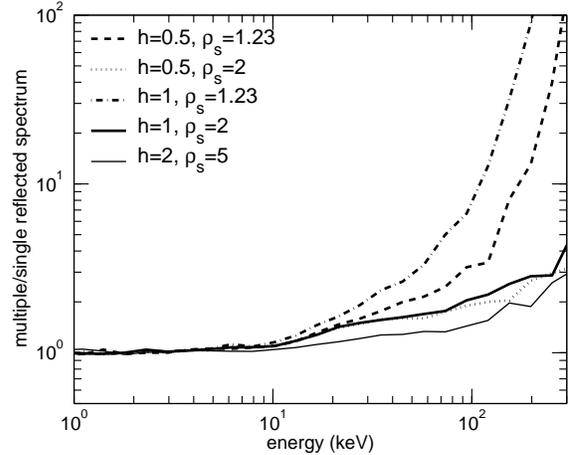,width=75mm}}
\centering
\caption{Effects of returning radiation on the reflected spectrum.
Ratio of reflected spectra calculated including multiple reflections
to single reflection only for the parameters shown on the figure. The inclination is $30^o$}
\label{fig:returnrad2}
\end{figure}

The first investigation on this effect is done by Cunningham (1976).
Multiple reflections occur when reflected photons are
deflected again toward the disc.
Ross et al.(2002) study multiple reflection spectra from ionized slabs in newtonian geometry.
These effects can be simply implemented into our numerical scheme, enabling us for the first time to evaluate accurately their influence.
We now compare our results with what is obtained when they are neglected.
 The result is that although the returning radiation has always a negligible effect on the total
reflected photon flux (less than 1 \%),  it can have a modest but
significant impact on the total
reflected energy flux, which is increased by up to 30 \% in the extreme case
 $\rho_{\rm s}=1.23$, $h\sim 1$
 (the effect is stronger when the source is closer to the black hole).
 Indeed, multiple reflection photons have on average a higher energy.
 In fact, when the irradiation is concentrated close to the black hole a significant fraction
 of the reflected photons returns to the disc.
 Such as for $h=1$ and $\rho_{\rm s}=1.23$, there are about 16 \% of reflected photons returning to the disc
 but there are only ~5 \%, for $h=5$ and $\rho_{\rm s}=5$.
 Below 10 keV most of them are absorbed while,
 at higher energy, more and more photons are reflected.
 The twice reflected  is therefore harder than the single reflected spectrum.
 Fig.~\ref{fig:returnrad1} compares reflected spectra
 calculated with and without returning radiation.
 Below 10 keV there is no significant difference and the returning radiation
 can safely be neglected. On the other hand, at higher energies,
 the two spectra differ significantly. Fig.~\ref{fig:returnrad2}
 shows the ratio of spectra computed with and without returning
 radiation for different values of $h$ and $\rho_{\rm s}$.
 The difference between the two spectra increases strongly with energy.
 In the most extreme cases, the returning radiation can enhance the reflected
 flux by more than one order of magnitude, above 100 keV.
We note that surprisingly, in our extreme limit $\rho_{s}=1.23$
 the effect is more important for $h=1$ than for $h=0.5$
 for which the illumination is concentrated closer to the black hole
 and the effects of returning radiation could be expected to be stronger.
This is because when the source gets so close to the black hole, more photons
 are trapped into the hole instead of intercepting the disc or reaching infinity.
Consequently both single and multiple reflections are reduced
 (it can be seen on Fig~\ref{fig:rfvspr}
  that the reflected flux is larger at $h=1$ than at $h=0.5$  by a factor of $\sim 2$).
At larger $\rho_{s}$ the effects of returning radiation
  are less dramatic and start being negligible
  for $h>2$ or/and $\rho_{s}>5$ (less than 10 \%  below 100 keV).
In conclusion, the effects of returning radiation are relatively weak except
  above 10 keV when the illumination is concentrated within a few gravitational radii.
These effects should be taken into account when fitting the spectra of
  extreme sources such as MGC 6-30-15
  with broad band missions such as BeppoSAX, INTEGRAL
  or the recently launched Suzaku.


\section{Connections with observations}\label{sec:observations}

In this section, we discuss our results in the context of current observations of AGN
and black hole binaries. In Sec.~\ref{sec:fitnewton}, we check the consequences of neglecting the GR effects on the resulting best fit parameters when fitting
observed spectra with a Newtonian reflection model.  In Sec.~\ref{sec:refvsprim} we discuss the predictions of the light bending model in the context
of recent monitoring campaigns showing, in several sources, the absence of a linear correlation between reflection (or iron line flux) and primary emission.
Finally, in Sec.~\ref{sec:flfl} we show that the light bending model produces a non-linear relation between fluxes measured in different energy bands that
we compare to the flux-flux relation observed in the narrow line Seyfert galaxy NGC4051.

\subsection{Fits with a Newtonian model}\label{sec:fitnewton}

\begin{table*}
\centering
\begin{tabular}{c|c|ccc|c|ccc}
  $\rho_s$ & $R_{GR}$ ($30^o$) & $\Gamma$ ($30^o$) & $R$ ($30^o$) & $\chi^2/dof$ ($30^o$) & $R_{GR}$ ($60^o$) & $\Gamma$ ($60^o$) & $R$ ($60^o$) & $\chi^2/dof$ ($60^o$) \\[3pt]
  \hline
  &&&&&&&\\
  2 s & 6.9290 & 2.5484 & 52.293 &  358/ 60 & 5.0553 & 2.4555 & 20.260 &  167/ 60\\[2pt]
  2 m & 8.3042 & 2.3500 & 35.850 &  210/ 60 & 6.0765 & 2.1809 & 11.562 &  83/ 60\\[6pt]
  5 s & 1.7605 & 1.8170 &  0.6400 &  81/ 60 & 2.0264 &  1.96093 & 1.4027 &  11/ 60\\[2pt]
  5 m & 1.9595 & 1.8732 & 1.1093 & 63/ 60 & 2.2319 & 1.9761 & 1.7323 & 13/ 60\\[6pt]
  10 s & 1.3532 & 1.9163 &  0.7715 &  39/ 60 & 1.6105 & 1.9710 & 1.1996 &  10/ 60\\[2pt]
  10 m & 1.4258 & 1.9347 & 0.9354 & 32/ 60 & 1.6918 & 1.9796 & 1.3394 & 13/60\\[6pt]
  20 s & 1.1641 & 1.9700 &  0.8956 &  26/ 60 & 1.3458 & 1.9754 & 0.9947 &  7/ 60\\[2pt]
  20 m & 1.1876 & 1.9636 & 0.8905 & 26/ 60 & 1.3628 & 1.9787 & 1.0368 & 7/ 60\\[4pt]
\end{tabular}
 \caption{Results from the spectral fits to the single(s), which do not take the returning radiation in consideration, and multiple(m) GR spectra for fixed source height, $h$=2, and different $\rho_s$ in
  energy range 2-30 keV for an inclination of $30^o$ and $60^o$ comparing to the $R$ calculated from GR as shown
  in second and sixth column, respectively. The energy cutoff and the observation inclination of PEXRAV model are frozen at the same values of GR.
  The  $\chi^2$ values are not statistically meaningful but based  on error bars fixed to 3 \% of the number of photons
   in each energy bin.}
\label{tab:pexrav}
\end{table*}

Actually, when dealing with real data, the amplitude of reflection $R$ is usually measured by spectral fits using a Newtonian reflection model.
 We therefore fitted our simulated GR spectra with the PEXRAV  model (Magdziarz \& Zdziarski 1995) under XSPEC.
 This procedure provides us with the estimate of $R$ that would be obtained if these GR spectra were observed and fit with this standard reflection model.
 The spectra calculated from PEXRAV correspond the NN with on-axis source. We fit the spectra in the energy range from 2 to 30 keV.
Since the formation of the iron line is included in our Monte-Carlo simulations and not in the PEXRAV model, the 6.8-8 keV energy band was ignored.
The results for $h$ = 2 with different $\rho_s$ are presented in Table~\ref{tab:pexrav}

\begin{figure}[th]
\centerline{\psfig{figure=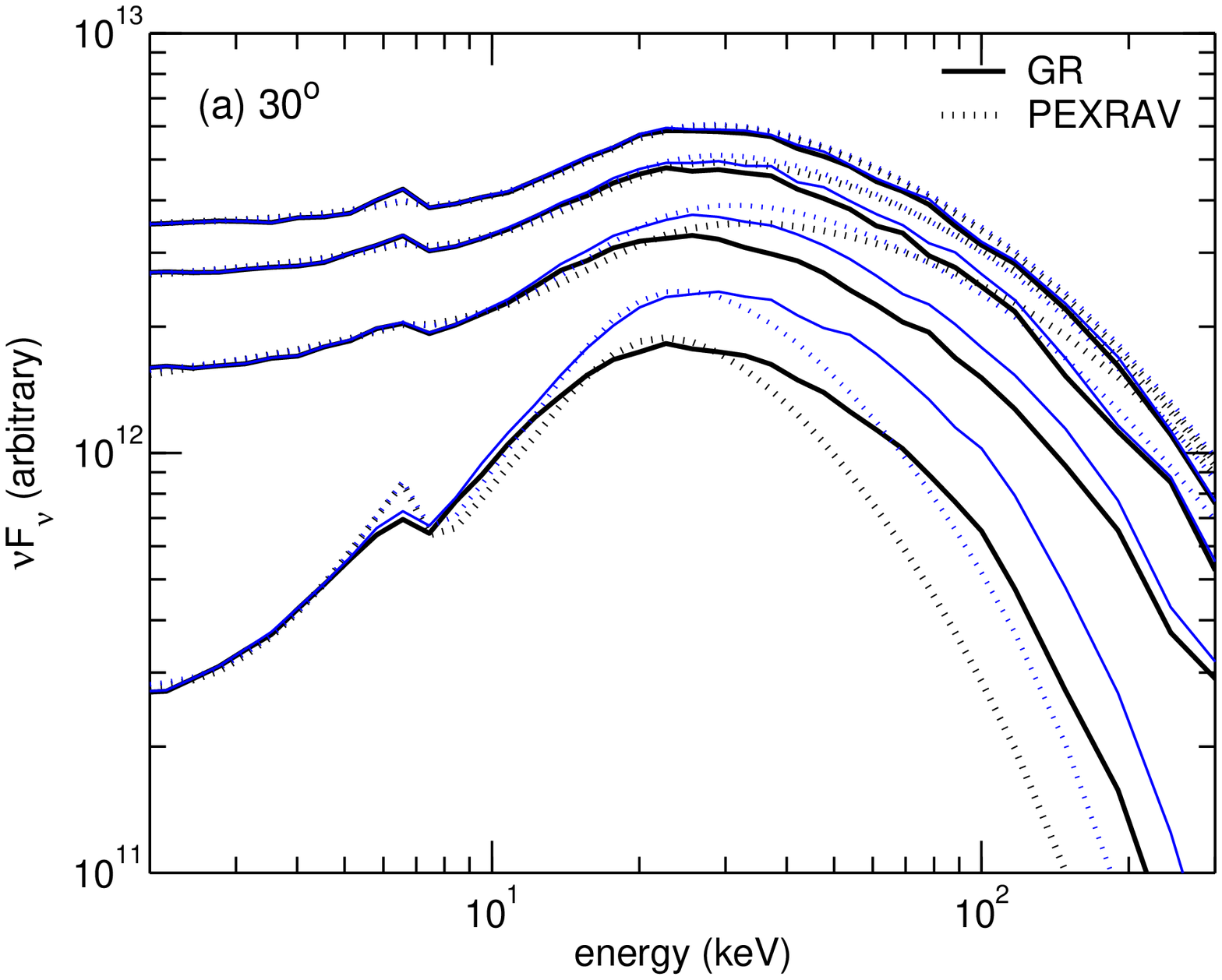,width=75mm,height=60mm}}
\centerline{\psfig{figure=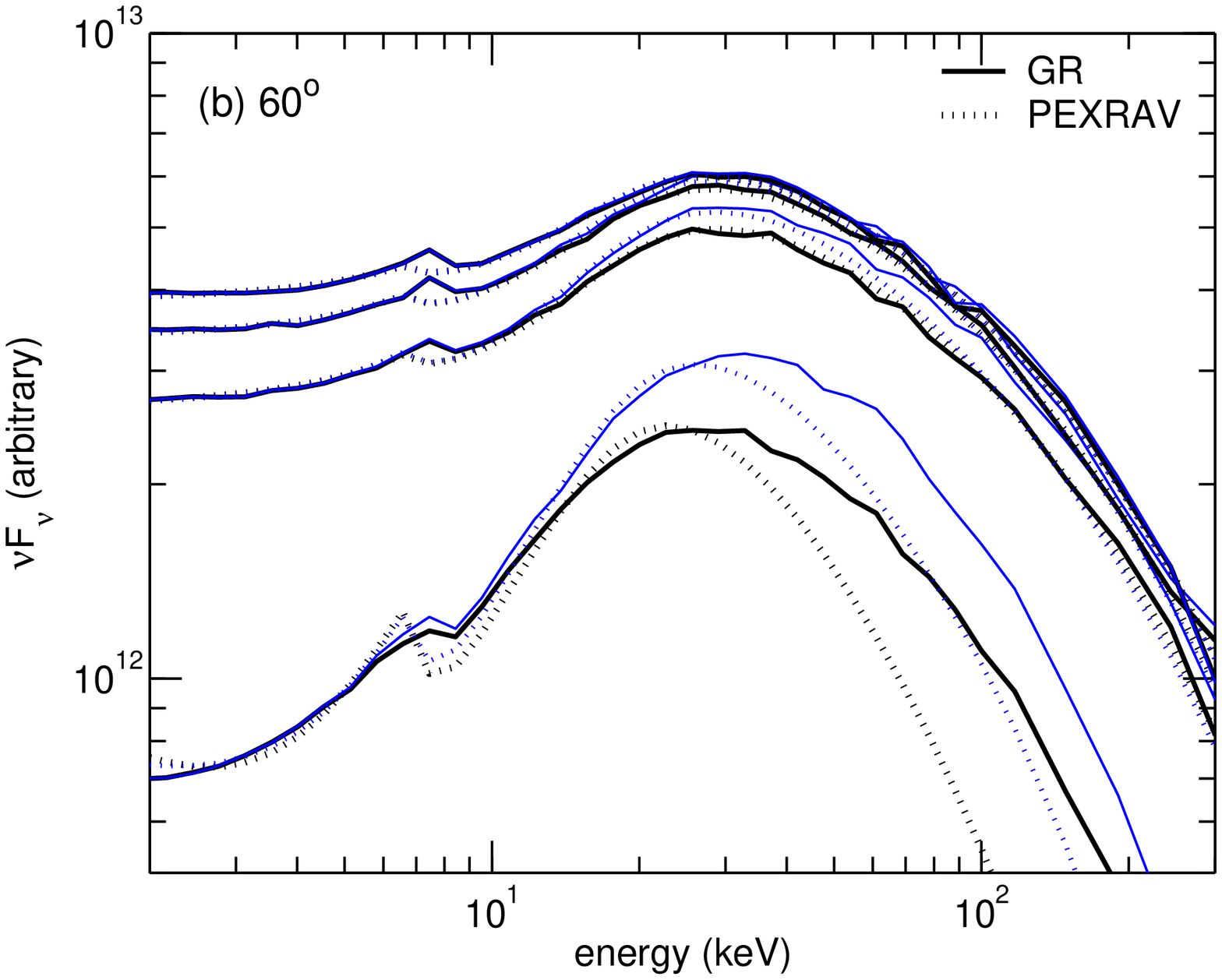,width=75mm,height=60mm}}
\centering
\caption{The spectra of GR (solid line) and the 2 -30 keV fits with PEXRAV model (dotted line),
 thick line for single reflection and thin line for multiple reflection,
   with $\rho_s =$ 2, 5, 10 and 20, from bottom to top respectively, for an inclination of $30^o$ (a) and $60^o$ (b).  }
\label{fig:fit}
\end{figure}

Although in most cases the 2-30 keV relativistic spectra are well described by the PEXRAV model (see Fig.~\ref{fig:fit}),
there are important differences between the values of $R$ estimated from spectral
fits and those determined theoretically from the RPF ratios.
These differences are due to the relativistic distortions of the reflected spectrum
and illustrate their  effects on the best-fit parameters
  obtained with a non-relativistic reflection model. Overall, fitting with PEXRAV
   tends to underestimate the reflection fraction.
 The discrepancies between the two estimates
 tend to be larger at small ring radii and inclination angles, since the GR
 spectral distortions are then more important.
 The $\chi^2$ values are not statistically meaningful since they depend on the arbitrary uncertainties that we had
 applied to the theoretical spectra, they are nonetheless interesting to compare the relative 'goodness' of the PEXRAV
 approximation to the different GR models. A comparison of the $\chi^2$ values obtained
 for the different fits also shows that the PEXRAV model
 gives a better representation of the simulated spectra at both large $\rho_{s}$
and large inclinations.

We also note that it is often impossible to find a reasonably good approximation of the GR
 spectra with PEXRAV over a broader energy range because the
GR reflected spectra are broader than the PEXRAV ones (see Fig.~\ref{fig:fit}).

\subsection{Reflection vs primary flux diagrams}\label{sec:refvsprim}

\begin{figure}[!h]
\centerline{\psfig{figure=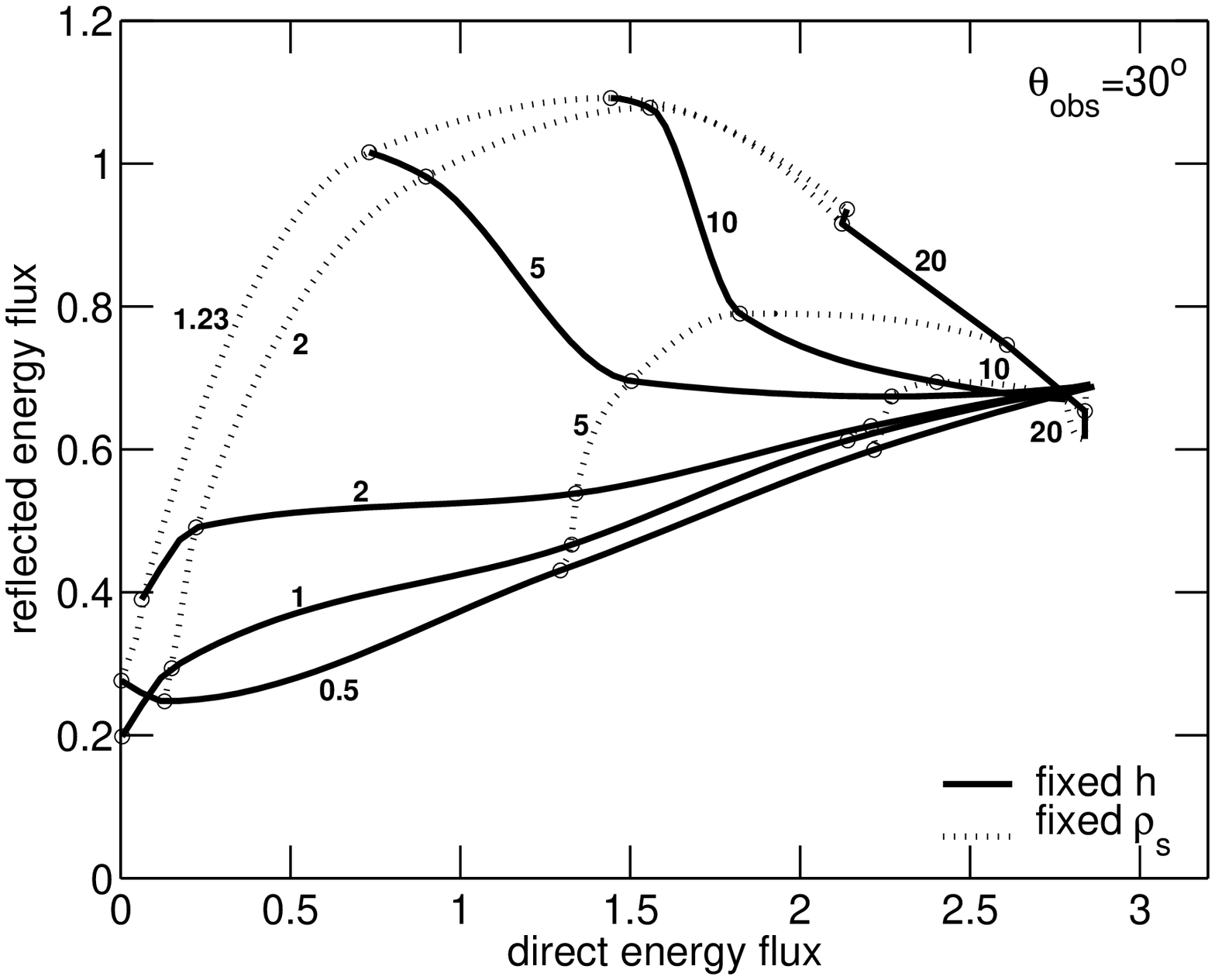,width=75mm, height=60mm}}
\centerline{\psfig{figure=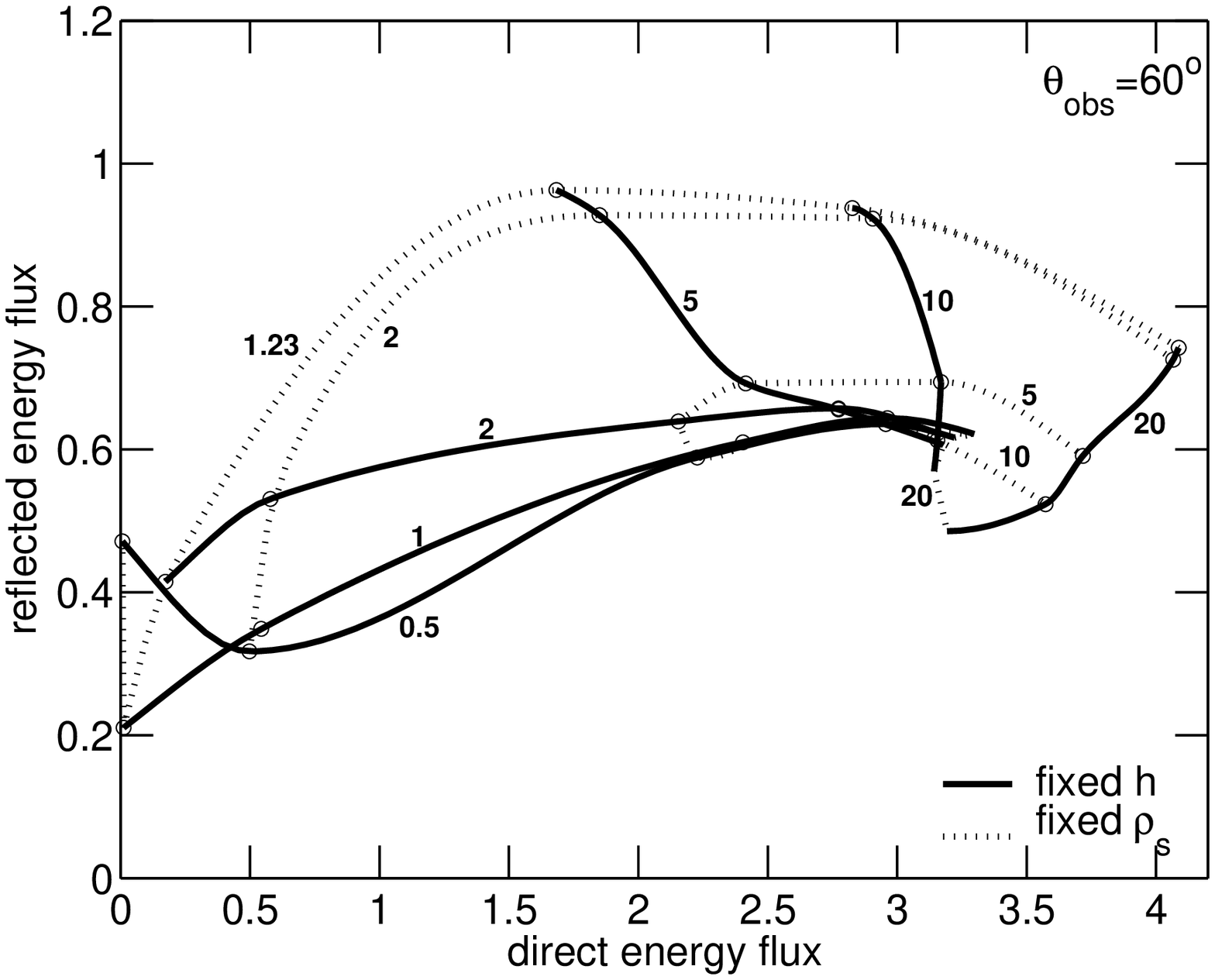,width=75mm, height=60mm}}
\centering
\caption{The total reflected energy flux as a function of the total direct energy flux,
both are in the energy range 1-30 keV, for an observation inclination $30^o$
 (top panel) and $60^o$ (bottom panel) with different source height and radius.
 The thin lines represent the fixed source radius and the thick lines indicate the same
  source height. These lines are interpolated from the results calculated from
  the model GR as shown at the intersection points. }
\label{fig:rfvspr}
\end{figure}

Fig~\ref{fig:rfvspr} shows the dependence of the reflected flux on
the primary emission when both $h$ and $\rho_{\rm s}$ are varied.
When the source height changes at constant radius and as long as $\rho_{s}\lesssim 5$,
 its track in this plane can be described according to three regimes:
i) at low fluxes (or low source height) the reflected and primary flux are correlated,
ii) at higher fluxes the reflection saturates at an almost constant value
while the primary can change by a factor larger than 2,
iii)  at even higher fluxes the reflection component is weakly anti-correlated
with the primary emission.
This behaviour is described in great details by Miniutti and Fabian (2004).
 As shown by these authors many properties of the variability of Seyfert galaxies
  and black hole binaries can be understood in terms of fluctuations of the source height.
  In particular, the monitoring of Seyfert galaxies indicates that the reflection
  flux  can be weakly variable when the primary emission
  changes dramatically (Papadakis et al. 2002; Markowitz et al. 2003).
  Moreover in at least two AGN, MGC 6-30-15 (Miniutti 2003; Reynolds et al. 2004)
   and NGC4051 (Ponti et al. 2006)  and one X-ray binary
  (XTE J1650-500, Rossi et al. 2004)  the reflection flux is correlated to the primary
  emission at low fluxes and saturates at higher fluxes, in qualitative agreement with
   the predictions of this model.

Fig~\ref{fig:rfvspr} enables us to investigate further the model parameter space.
It shows that if the radius is larger than $\sim  5$ the variability induced
 by change in the height is much too weak ($< 2$) to account for the variability observed in most accreting black holes.
Therefore, if change in the source height, in the context of the light bending model,
 is to be responsible for the variability properties of MCG 6-30-15, NGC 4051 and
 XTE J1650-500, the illumination has to be concentrated within a few gravitational radii
 in these sources.

Let now consider the effects of changes in the source radius at constant height.
At small source heights ($h\lesssim  5$), the overall trend is that the reflection and
 primary emission are weakly correlated: the reflected flux changes by at most 50 \%
 when the primary flux increases by more than one order of magnitude which might be in qualitative agreement
 with some observations but is inconsistent with the strong non-linear
 correlation observed  for instance in the low state of NGC4051.

At higher source heights, the reflected and primary flux become
 anti-correlated, which is not observed.
The slope of the anti-correlation increases with $h$.
At $h\sim 10$ we could observe large variations of the reflection
  component at constant primary flux.
At even larger $h$ ($\gtrsim 10$) the two components become correlated again
  but with a small amplitude of variability because of the vanishing light bending effects.

These results show that if the light bending model is to be the correct
  interpretation of the observations,
  the driver of the variability should be $h$
  while the source radius has to be nearly constant and reasonably small ($\lesssim 5$).

\subsection{Flux-flux diagrams: application to NGC4051}\label{sec:flfl}

\begin{figure*}[th]
\centerline{\psfig{figure=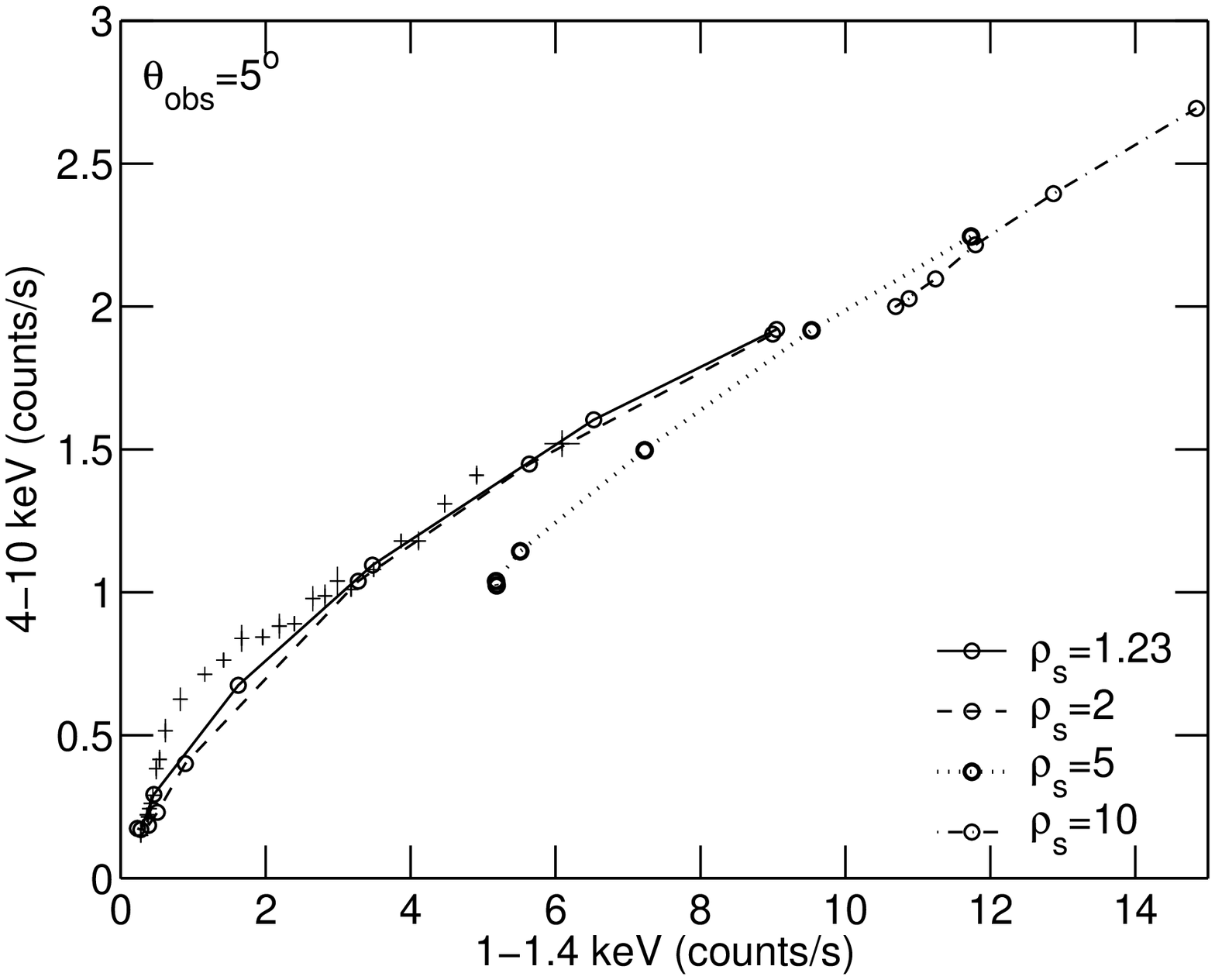,width=80mm}\psfig{figure=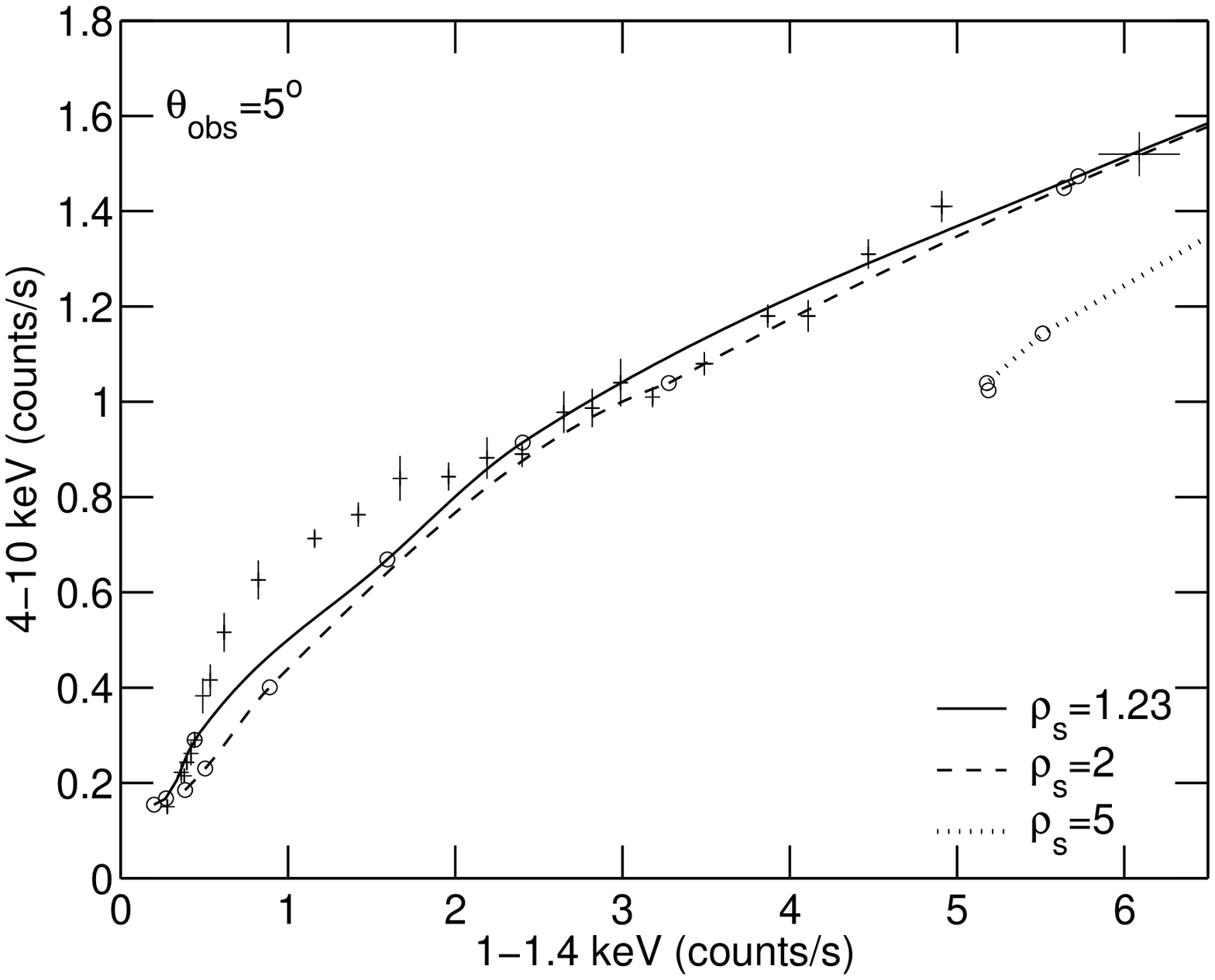,width=80mm}}
\centering
\caption{Flux-flux correlations obtained when varying $h$ from 0.5 to 20 at
 constant $\rho_{\rm s}$. The lines represent an interpolation between
 the simulated results (shown by the circles) for an inclination of $5^o$ and various value of $\rho_{s}$.
The crosses are the NGC4051 data from Ponti et al. (2006). The right panel
l is an enlarged version of the left one showing
 the region where model and data overlap. In all simulations the inclination is of $5^o$.}
 \label{fig:ngc4051rho}
\end{figure*}

 A direct and quantitative comparison of our calculations with data is difficult.
 Besides the nuclear reflection and primary emission that we are attempting to model here,
 the observed spectra of accreting black holes can be affected
 by a number of features, such as absorption, emission lines from distant
 photo-ionised plasma or distant reflection. Fitting observational data using our model
 is therefore deferred to further work. On the other hand,
 using the published estimates of the reflected and
  primary emissions of observed sources for comparisons with our simulated reflection-primary flux diagrams,
  might be inaccurate because these estimates are model
  dependent. For instance Ponti et al. (2006) performed a spectro-temporal analysis
  of two XMM-Newton  observations of the Narrow Line Seyfert 1, NGC 4051.
  They produced a reflected versus primary flux diagram showing a correlation at low
   fluxes and a saturation of the reflected flux at higher luminosities, in qualitative
   agreement with what is obtained in the light bending model when varying $h$ at constant $\rho_{\rm s}$.
   However in their spectral analysis they used a ionised reflection model and found that
    the data require the disc to be mildly ionised.
  In such conditions, the disc albedo is much larger than in the neutral case
  (see e.g. Malzac, Dumont \& Mouchet 2005)
  and ionised reflection produces a strong soft excess below 1 keV which is not
   present if the disc is neutral.
  Therefore the exact numerical values of the integrated reflected flux obtained from
  their spectral fitting are not comparable with the results of our neutral model.
To compare our results with data in a way that is less sensitive to ionisation
  and other spectral complications we will use flux-flux diagrams.
These flux-flux plots reveal how the fluxes in two different energy bands relate
  to each other. They are often used to study the variability of AGNs and X-ray binaries
  independently of any model (see Taylor, Uttley \& McHardy 2003; Uttley et al. 2004).
In Fig~\ref{fig:ngc4051rho} we compare the flux-flux relation predicted by the light bending model
 (assuming no intrinsic source fluctuations)
 to the data reported by Ponti et al. (2006).  To produce this figure, we simulated
light bending model spectra for the spectral index of primary emission
observed in NGC4051 ($\Gamma=2.3$) and various source heights and radii.
We computed the predicted photon flux in the two energy bands (1-1.4 and 4-10 keV)
used by Ponti et al. (2006). In these two bands, the effects of ionisation
of the reflection component on the total flux are weak.

As the data of Ponti et al.(2006) are given in counts, they cannot be compared directly
with the photon fluxes.
The overall slope of the count-rate correlation
also depends on the ratio of the detection efficiencies of XMM-Newton in these two bands.
We estimated this ratio by convoluting simple power-law spectra with
the EPIC pn response matrix used by Ponti et al. (2006).
 We found that this efficiency ratio is not sensitive to
the shape of the model spectrum.
 It varies by  less than 10\%   for power-law indices $\Gamma$ in the range $1.3-2.5$.
 This range of $\Gamma$ is broader than the observed variations of the 2-10 keV spectral index in
 NGC4051.  We therefore neglected the fluctuations of the efficiency ratio
  and used its average value to correct the model photon
fluxes. Then, we had to fix the intrinsic ring source luminosity, which controls
simultaneously the minimum and maximum  achievable fluxes in both energy bands.
 It turns out that this parameter is very constrained by the observations.
 If it is too high, the minimum measured flux cannot be reproduced by the model.
 If, on the other hand, it is too low, the model points fall well below the observed
 correlation whatever $h$ and $\rho_{\rm s}$. The only solution is to set
 the intrinsic ring source luminosity so that the point obtained for the extreme
 case $\rho_{s}=1.23$ and $h=0.5$, matches that of the minimum observed fluxes.
 Then it appears that the overall shape and amplitude of the correlation
  is qualitatively reproduced by the model for $h$ varying approximately
  between 0.5 and 10 and $\rho_{s}\lesssim 2$ as shown in Fig~\ref{fig:ngc4051rho}.
  Of course, the detailed shape of the observed correlation
 is not perfectly reproduced by the model.
  Nonetheless, if one considers the simplicity of the model, as well as the
   many complications that could
  affect the shape and normalisation of the flux-flux correlation
   (fluctuation of primary spectral slope or the intrinsic source luminosity,
   effects of ionisation of the disc...) the agreement
   we obtain is remarkable.

\begin{figure}[h]
\centerline{\psfig{figure=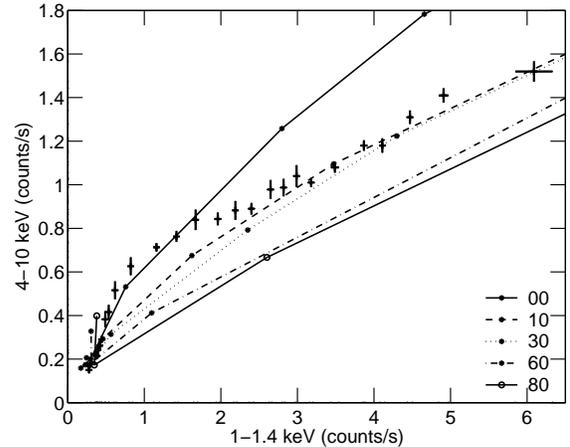,width=75mm}}
\centering
\caption{Flux-flux correlations obtained when varying $h$ from 0.5 to 20 at
 $\rho_{\rm s}=1.23$ and various inclination angles.
  The lines represent an interpolation between the simulated results (shown by the circles).
The crosses are the NGC4051 data from Ponti et al. (2006).}
 \label{fig:ngc4051angle}
\end{figure}

  Naturally, the correlation predicted by the model depends on the inclination angle.
  Regarding NGC4051, we find a better agreement with the observed correlation
  for smaller inclinations (see Fig~\ref{fig:ngc4051angle}), in agreement with
 the Seyfert unification scheme (Antonucci 1993).
  Indeed, at large inclination the model correlation is more linear and
  the bending at low fluxes of the observed correlation is not well reproduced.
  On the other hand, if the source is seen exactly pole-on the minimum attainable
   flux in the 4-10 keV band is reduced by a factor of about 2, which allows the
correlation obtained
for $\rho_{s}=1.23$ to be closer to the data at low fluxes and largely above
at higher fluxes, leaving room  for a somewhat broader range in  $\rho_{s}$
(up to $\rho_{s}\sim 4$).
In this case, $h$ and $\rho_{\rm s}$
would have to vary simultaneously along the observed flux-flux correlation track.
The emitting region is therefore confined within a few gravitational radius.

We note that for the quite extreme parameters that we infer the broad iron line is so smeared that it is difficult to detect
(see Fig.~\ref{fig:returnrad1}). Pounds et al. (2004) fit the spectra produced
from the same XMM-Newton data that we used and found that the curvature of the spectrum around the line energy can be explained
 by a partial covering model without any broad iron line (but see Ponti et al. 2006).
 In contrast, other studies present evidences for a broad Fe~K$_{\alpha}$ line in NGC4051
  (Guainazzi et al. 1996; Uttley et al. 2003).
Fitting the XMM-Newton spectra with a relativistically blurred reflection model requires
 the illumination to be extremely concentrated in the innermost parts of the accretion disc
 (Ponti et al. 2006 derive a disc emissivity index $q\sim 5$) in agreement with the numbers
 inferred here.

\section{Conclusion}\label{sec:conclusion}

Considering an accretion disc illuminated by a ring source of hard X-rays,
we have studied the dependence of the observed primary emission
and reflection component on the radius of the ring source, its height above the disc
 and the inclination of the observer's line of sight.
We confirm the results of Miniutti et al. (2004) showing that the general relativistic
 effects can lead to strongly reflection dominated spectra
 when the primary source is very close to the black hole.
 Fitting these GR models with the newtonian reflection model PEXRAV
  leads to underestimate the reflection fraction $R$.
Moreover varying the height or radius of the ring source can produce important
 variations of the observed primary luminosity with little variations
of the reflected flux.
Therefore fluctuations of the height or radius of the ring source
can lead to variability modes in which the primary
and reflected components are apparently decoupled, as observed in several sources (Papadakis et al. 2002; Markowitz et al. 2003;
Miniutti 2003; Reynolds et al. 2004; Ponti et al. 2006; Rossi et al. 2004).
We have shown that the light bending model can produce a non-linear flux-flux relation
that is similar to that observed in several sources.
In particular we compared our model with the data of NGC4051, and
 found an acceptable agreement. This rough
  comparison suggests a low inclination angle ($<20^o$), an
  illuminating source at a radius $\lesssim$ 3
  and a height varying from almost 0 up to about 10 gravitational radii.

Regarding the angular distribution of the radiation,
 we have found some important qualitative differences with respect to the Newtonian case.
 In particular, the reflected flux at larger inclination is relatively stronger than
 in the Newtonian model.
 If the ring source radius is zero (on-axis model),
 the total reflected flux as well as the reflection coefficient $R$
is larger at large inclination.
In the off-axis model the total reflected flux decreases at larger inclinations but less sharply
 than in the Newtonian model, as a consequence $R$ can be larger at large inclinations.

We also found that the effects of returning radiation on the amplitude and shape
of the reflection component can be safely neglected
 below 10 keV, but at higher energies the reflected flux can be enhanced by several
  orders of magnitude. This spectral distortion should be taken into account when fitting
  broadband spectra of accreting black holes which emission comes from regions very close
  from a black hole.

\begin{acknowledgements}
      We thank P.O. Petrucci for help in checking the results of our numerical scheme,
      Andrea Martocchia and Giovanni Miniutti for useful discussions and
      important suggestions, Gabriele Ponti for providing us with the NGC4051
      data before publication.
      JM acknowledges support from the National Science Foundation under
      Grant No. PHY99-07949.
\end{acknowledgements}

\appendix
\renewcommand{\theequation}{A\arabic{equation}}  
\setcounter{equation}{0}  

\section{Constants of motion}
The Kerr black hole  metric is described in the Boyer-Lindquist spherical coordinates (BLC) as,
\begin{align}
{ds}^{2}= &- (1 - \frac {2\,M\,r}{\Sigma })\,dt^{2} - \frac {4\,M\,a\,r\,sin^{2}\,\theta \,dt\,d\phi}{\Sigma } \notag \\ &+ \frac {A\,sin^{2}\,\theta\,d\phi ^{2}}{\Sigma } + \frac {\Sigma \,dr^{2}}{\Delta} + \Sigma \,d\theta^{2}
\end{align}
where, ($G=c=1$),
$$ \quad \Delta=r^{2} + a^{2}-2Mr,\quad A=(r^{2} + a^{2})^{2} - a^{2}\Delta sin^{2}\,\theta,$$
$$\quad \Sigma=r^{2} + a^{2}cos\,\theta,\quad \omega=\frac{2Mar}{A}.$$
we can write the metric appropriate to a stationary axisymmetric space-time in the form (Chandrasekhar 1983)
\begin{equation}
ds^2\,=\,-e^{2\nu}\,dt^2\,+\,e^{2\psi}(d\phi-\omega dt)^2\,+\,e^{2\mu_1}dr^2\,+\,e^{2\mu_2}d\theta^2
\end{equation}
where
$${e^{2\mu_1}}=\frac {\Sigma }{\Delta },\;\;\;{e^{2\mu_2}}=\Sigma, \;\;\;{e^{2\psi }}=\frac {A\,sin^{2}\,\theta}{\Sigma },\;\;\;{e^{2\nu }}=\frac {\Sigma \,\Delta }{A}.$$
In this form, we can define a tetrad frame comoving with emitter, $dr/dt=0$, $d\theta/dt=0$ and $d\phi/dt=\Omega$, by the contravariant basis vector, $e_{(a)}\!^i$:
\begin{align}
e_{(t)}\!^i = (&\frac{e^{-\nu}}{\sqrt{1-e^{2\psi}(\Omega-\omega)^2/e^{2\nu}}}, \frac{e^{-\nu}\Omega}{\sqrt{1-e^{2\psi}(\Omega-\omega)^2/e^{2\nu}}}, 0, 0 ), \notag \\
e_{(\phi)}\!^i = (&\frac{e^{\psi-\nu}(\Omega-\omega)}{\sqrt{e^{2\nu}-e^{2\psi}\Omega^2+2e^{2\psi}\Omega\omega-\omega^2 e^{2\psi}}}, \notag \\ &\frac {e^{-\nu-\psi}(e^{2\nu}-\omega^{2}{e^{2\psi}}+e^{2\psi}\Omega\omega)}{\sqrt {e^{2\nu}-e^{2\psi}\Omega^2+2e^{2\psi}\Omega\omega-{\omega}^{2}{e^{2\psi}}}}, 0, 0 ), \notag \\
e_{(r)}\!^i = (& 0, 0, e^{-\mu_1} ,0 ), \notag \\
e_{(\theta)}\!^i = (& 0, 0, 0, e^{-\mu_2} ).
\end{align}
We use the tetrad frame's basis vector to express the relation between local photon's direction and the constants of motion as
\begin{align}
(E_0,\,E_0\,sin\,\alpha\,cos\,\beta,\,E_0\,cos\,\alpha,\,E_0\,sin\,\alpha\,sin\,\beta) &= \notag \\ e_{(a)}\!^i \;(E,L,&P_r,P_\theta)
\end{align}
where $E_0$ is the magnitude of photon momentum at local rest frame when $c = 1.$ $\alpha$ is polar angle direction and $\beta$ is azimuthal angle direction in the spherical local emitter's rest frame using radial direction as axis. For the radial direction toward the centre of black hole, $\alpha=\pi.$ For directions paralleling accretion disc, $\alpha=\pi/2$, and countering or following black hole rotation, $\beta=0$ or $\beta=\pi$, respectively. $P_r$ and $P_\theta$ are the potential momentums in BLC in $r$ and $\theta$ direction, respectively. We obtain the constants of motion:
\begin{align}
L &= \frac{-E_0((\Omega-\omega){e^{2\psi}}+sin\,\alpha\,sin\,\beta\,e^{\psi+\nu})}{U_0(-{e^{2\nu}}+\omega^2{e^{2\psi}}-2\Omega\omega{e^{2\psi}}+\Omega^2{e^{2\psi}})}\label{eq:contantsofmotion1}\\
E &= \frac{E_0((-{e^{2\nu}}+\omega^2{e^{2\psi}}-\omega\Omega{e^{2\psi}})-\Omega sin\,\alpha\,sin\,\beta\,e^{\psi+\nu})}{U_0(-{e^{2\nu}}+\omega^2{e^{2\psi}}-2\Omega\omega{e^{2\psi}}+\Omega^2{e^{2\psi}})}\label{eq:contantsofmotion2}\\
K &=(E_0sin\,\alpha\,cos\,\beta\,e^{\mu_2})^2-cos^2\,\theta((aE)^2-(L/sin\,\theta)^2)
\label{eq:contantsofmotion3}
\end{align}

where
$$U_0 = \frac{dt}{ds} = \frac{1}{e^\nu\sqrt{1-V^2}} \quad\rm and\quad V^2 = {e^{2\psi-2\nu}}(\Omega-\omega)^2. $$

\end{document}